\documentclass[a4paper]{article}
\usepackage[utf8]{inputenc}
\usepackage[T1,T2A]{fontenc}
\usepackage[
unicode,
pdfauthor={金炯錄},
pdftitle={L∞‐algebraic extensions of non‐Lorentzian kinematical Lie algebras, gravities, and brane couplings},
pdfsubject={hep-th,gr-qc,math-ph},
breaklinks=true,
final
]{hyperref}
\usepackage{amsmath,amssymb,amsthm,orcidlink,booktabs,cleveref,pdflscape}
\usepackage[noadjust,nosort]{cite}
\usepackage{CJKutf8}
\usepackage[final]{microtype}
\usepackage[latin,russian,vietnamese,german,italian,british]{babel}
\usepackage[osf,theoremfont,largesc,p]{newpxtext}
\usepackage[varbb,slantedGreek]{newpxmath}
\usepackage[artemisia]{textgreek}
\usepackage{xcolor}
\usepackage{tikz-cd}
\usetikzlibrary{babel}

\newtheorem{proposition}{Proposition}
\theoremstyle{definition}
\newtheorem{definition}{Definition}
\renewcommand\delta\deltaup
\renewcommand\varepsilon\varepsilonup
\renewcommand\Omega\Omegaup
\renewcommand\Re{\operatorname{Re}}
\renewcommand\Im{\operatorname{Im}}

\begin{document}
\title{\(L_\infty\)-algebraic extensions of non-Lorentzian kinematical Lie algebras, gravities, and brane couplings}
\author{
Hyungrok Kim~(\begin{CJK*}{UTF8}{bsmi}金炯錄\end{CJK*})\textsuperscript{\orcidlink{0000-0001-7909-4510}}\footnote{Centre for Mathematics and Theoretical Physics Research, Department of Physics, Astronomy and Mathematics, University of Hertfordshire, Hatfield, Hertfordshire \textsc{al10 9ab}, United Kingdom}
\\ [1em]
\texttt{\href{mailto:h.kim2@herts.ac.uk}{h.kim2@herts.ac.uk}}
}
\maketitle
\begin{abstract}
The Newtonian limit of Newton--Cartan gravity relies crucially on the Lie-algebraic central extension to the Galilean algebra, namely the Bargmann algebra. Lie-algebraic central extensions naturally generalise to \(L_\infty\)-algebraic central extensions, which in turn classify branes in superstring theory via the brane bouquet.
This paper classifies all \(L_\infty\)-algebraic central extensions of all kinematical Lie algebras that do not depend on the spatial rotation generators as well as all iterated central extensions thereof (for codimensions \(\le3\)). The Bargmann central extension of the Galilean algebra then appears as merely one term in a sequence of \(L_\infty\)-algebraic central extensions in each degree; a similar situation obtains for the Newton--Hooke algebra and the static algebra, but not for the Carrollian algebra nor those kinematical Lie algebras that are not Wigner--İnönü deformations of a simple algebra.

The sequence of \(L_\infty\)-algebraic central extensions in each degree then corresponds to a tower of \(p\)-form fields. After imposing conventional constraints, the zero-form field provides absolute time, and the higher-form fields are certain wedge products of the field strengths of the one-form (Bargmann) gravitational field. These then provide natural \((p-1)\)-brane couplings to the corresponding non-Lorentzian gravities, which are found to produce velocity-dependent gravitational effects in the presence of torsion.
The \(L_\infty\)-algebraic cocycles also provide Wess--Zumino--Witten terms for the \((p-1)\)-brane action, which require the introduction of doubled spatial coordinates that are reminiscent of double field theory, but which (in some cases at least, and given appropriate kinetic terms) do not result in doubled physics.
\end{abstract}
\tableofcontents

\section{Introduction}
Field theories and string theories with non-Lorentzian spacetime symmetries have received much attention in recent years (as reviewed in \cite{Hartong:2022lsy,Bergshoeff:2022eog,Bergshoeff:2022iyb,Figueroa-OFarrill:2022nui}), arising in various contexts such as flat-space and non-Lorentzian holography \cite{Barnich:2009se,Bagchi:2010zz,Bagchi:2012cy,Christensen:2013lma,Christensen:2013rfa,Blair:2025nno}, the physics of null hypersurfaces \cite{Donnay:2019jiz,Duval:2014uva}, post-Newtonian corrections \cite{Hansen:2019svu,Hansen:2020wqw}, and fractons \cite{Nandkishore:2018sel,Pretko:2020cko}.
The spacetime symmetries also determine gravitational physics: Einstein gravity may be obtained by gauging Poincaré symmetry and then constraining certain components of the curvature (namely, the torsion) to vanish \cite{Freedman:2012zz}; Newtonian gravity may be similarly obtained by gauging a central extension of Galilean symmetry and constraining certain components of the curvature to vanish \cite{Andringa:2010it,Andringa:2013mma}.\footnote{Mre generally, one may gauge certain \(L_\infty\)-algebras to obtain gravitational theories \cite{Borsten:2024alh,Borsten:2024pfz}.}
In this procedure, it is crucial to take the (maximal) central extension of Galilean symmetry: the Newtonian gravitational potential originates from the central extension, and it is this vector field to which particles couple.

Central extensions of a Lie algebra \(\mathfrak g\) are classified by the second Lie algebra cohomology \(\operatorname H^2(\mathfrak g)\). There is nothing special about the number two; in general, the \(k\)th Lie algebra cohomology \(\operatorname H^k(\mathfrak g)\) instead classifies central extensions of \(\mathfrak g\) regarded as an \(L_\infty\)-algebra, which is the natural homotopy-theoretic generalisation of Lie algebras \cite{Jurco:2018sby,Kraft:2022efy,Lada:1992wc}.

Such \(L_\infty\)-algebraic central extensions of gauge symmetries arise in two related contexts.
\begin{enumerate}
\item In higher gauge theory \cite{Borsten:2024gox}, it is very natural to take central extensions of the gauge Lie algebra \(\mathfrak g\), and the extension given by \(\operatorname H^k(\mathfrak g)\) corresponds to introducing a \((k-1)\)-form gauge potential; abstractly, the gauge group is then generalised to an \(\infty\)-Lie group. For example, for every simple Lie algebra \(\mathfrak g\), one always has \(\operatorname H^3(\mathfrak g)\), corresponding to the \(L_\infty\)-algebra \(\mathfrak{string}(\mathfrak g)\) (the so-called string algebra), whose underlying graded vector space is \(\mathfrak g\oplus\mathbb R[1]\) with a nontrivial ternary bracket \(\mu_3\colon\mathfrak g\otimes\mathfrak g\otimes\mathfrak g\to\mathbb R[1]\) given by the 3-cocycle; this gauge algebra naturally yields six-dimensional theories with features similar to that of M5-branes \cite{Samann:2017sxo,Samann:2019eei,Rist:2020uaa}.
\item In string theory, certain central extensions of the super-Poincaré algebras (specifically, those that only depend on the translation and supertranslation generators) classify possible branes as well as determining which branes can end on which, a construction known as the \emph{brane bouquet} \cite{Fiorenza:2013nha,Fiorenza:2016ypo,Fiorenza:2016oki,Huerta:2017utu,Sati:2018tvj}.
Concretely, such \(L_\infty\)-algebraic central extensions give rise to Wess--Zumino--Witten terms in the Green--Schwartz formulation of brane actions that ensure \(\kappa\)-symmetry.
\end{enumerate}
This raises the following question: what \(L_\infty\)-algebraic central extensions do non-Lorentzian spacetime symmetry algebras admit, and what kinds of gravitational theories and branes do they yield? In this paper we take first steps towards answering these questions.

First, in
\cref{prop:jfree_cohomology_nonexceptional,prop:jfree_cohomology_exceptional}, we classify all \(L_\infty\)-algebraic central extensions of all kinematical Lie algebras
corresponding to cocycles that only involve boosts and spacetime translations, but not the \(\mathfrak o(d)\) spatial rotation generators \(\mathtt j_{ij}\) of the kinematical Lie algebra. (This restriction is motivated by the fact that rotation-generator-containing cocycles are ignored in the brane bouquet as well as technical convenience.) We find that the Bargmann central extension of the Galilean algebra (corresponding to Newton--Cartan gravity) appears as merely one term in a sequence of \(L_\infty\)-algebraic central extensions in each degree; a similar situation obtains for every non-simple, non-Poincaré kinematical Lie algebra that can be obtained as a Wigner--İnönü contraction of a simple Lie algebra except for the Carrollian algebra (namely, the static kinematical Lie algebra and the two Newton--Hooke algebras). For the Carrollian algebra in \((d+1)\)-dimensional spacetime, we find cocycles in degrees \(d\), \(d+1\), and \(2d+1\). The picture is more complicated for the exceptional kinematical Lie algebras in \(d\in\{2,3\}\) (see \cref{prop:jfree_cohomology_exceptional}).

The brane bouquet furthermore includes \emph{iterated} central extensions that encode branes ending on other branes; for this, we classify all iterated \(L_\infty\)-algebraic central extensions of kinematical Lie algebras that do not involve rotation generators and only involve branes of codimension \(\ge3\) in spacetime in \cref{prop:iterated_classification}. (The assumptions are motivated by qualitatively new physics entering in codimensions \(\le2\) as well as technical convenience.) The only nontrivial case is that of the Galilean, static, and Newton--Hooke algebras, for which we give an algebraic description of the possible iterated central extensions.

When gauged, an \(L_\infty\)-algebraic central extension given by a cocycle of degree \(p+1\) corresponds to the introduction of a \(p\)-form field in the gravitational theory; this generalises how the Bargmann extension of the Galilean algebra (given by a cocycle of degree two) corresponds to the introduction of the one-form field that encodes the Newtonian gravitational field. For the maximal \(L_\infty\)-algebraic central extension of the Galilean algebra, we thus obtain a series of differential form fields of every degree.
After imposing certain conventional constraints, we find that the zero-form field measures absolute time, the higher-form fields are certain wedge products of the field strengths of the one-form (Bargmann) gravitational field.

The central extensions in turn produce natural couplings to brane worldvolume actions, such as those that might appear in nonrelativistic string theories (reviewed in \cite{Oling:2022fft}).
A \((p-1)\)-brane naturally couples to a \(p\)-form field. In Newton--Cartan gravity, where the gravitational field is described by a one-form field, particles (zero-branes) naturally couple to gravity but higher-dimensional branes do not, in contrast to gravities based on a metric tensor such as Einstein gravity, to which branes of every dimension couples naturally. When one takes the maximal \(L_\infty\)-algebraic central extension of the Galilean algebra, we see that branes of every dimension can couple to a gravitational field; we show that, assuming certain torsion constraints, this produces velocity-dependent gravitational couplings on \(p\)-branes with \(p\ge1\).

Furthermore, following the paradigm of the brane bouquet, the \(L_\infty\)-algebraic cocycles also provide Wess--Zumino--Witten terms for brane actions. However, a complication arises: since the cocycles involve both spacetime translations as well as boosts, and since boosts and spatial translations are on an equal footing in the definition of kinematical Lie algebras, we see that the embedding maps of the brane must involve \emph{doubled} spatial coordinates (but the time coordinate remains not doubled) --- a situation reminiscent of doubled sigma models \cite{Tseytlin:1990va,Tseytlin:1990nb,Hull:2004in,Berman:2007xn,Copland:2011wx,DeAngelis:2013wba} that capture T-duality. However, while T-duality has been explored in non-Lorentzian string theory \cite{Bergshoeff:2018yvt,Kluson:2018vfd,Kluson:2019xuo,Gomis:2020izd} (cf.\ the review \cite{Oling:2022fft}), the relation to T-duality of these doubled spatial coordinates is not clear. If one only puts kinetic terms for half of the coordinates, then in some cases the other half can be integrated out as auxiliary fields, reminiscent of the choice of a solution to the section constraint in double field theory (reviewed in \cite{Aldazabal:2013sca,Hohm:2013bwa,Berman:2020tqn}).

\paragraph{Limitations and future directions.}
This paper only classifies those Lie-algebra cocycles that do not involve the \(\mathfrak o(d)\)-valued generator \(\mathtt j_{ij}\) corresponding to spatial rotation.
If one allows these, then there are nontrivial cocycles even for Poincaré and (anti-)de~Sitter algebras,
such as that corresponding to the string algebra.
While the corresponding gravities with \(p\)-form fields can be straightforwardly written down, an interpretation of the corresponding Wess--Zumino terms for the brane bouquet is even less obvious than in those in the present paper.

A natural further step would be to generalise the present discussion to incorporate central extensions of kinematical \emph{super}algebras, whose gauging would yield non-Lorentzian supergravities.
It is well known that gauging super-Poincaré \cite{Freedman:2012zz} and super-Bargmann \cite{Andringa:2013mma,Bergshoeff:2022iyb} algebras yield corresponding supergravities,
and the language of \(L_\infty\)-algebras extends naturally to \(L_\infty\)-superalgebras (graded by \(\mathbb Z\times\mathbb Z_2\) rather than \(\mathbb Z\)), not to mention that the brane bouquet was originally formulated for superstrings.

The present discussion does not attempt to incorporate adjustments \cite{Borsten:2024gox} of the \(L_\infty\)-algebras.
In higher gauge theory, if one wishes to avoid constraining any components of the curvature, the field strengths must be of a non-canonical form not specified by the gauge \(L_\infty\)-algebra alone \cite{Sati:2008eg,Sati:2009ic,Samann:2019eei}; the additional datum specifying the form of the field strengths is called an adjustment \cite{Rist:2022hci}.
In the present context, since we wish to constrain many components of the field strengths regardless, it is not clear that an adjustment is necessary. Furthermore, to construct adjustments it is technically convenient to work with strict models of \(L_\infty\)-algebras (i.e.\ those that are merely differential graded Lie algebras); while strict models are always known to exist (via the cobar construction), they are usually inconveniently large and difficult to deal with.
Nevertheless, it may be interesting to consider whether the (iterated) central extensions of \(L_\infty\)-algebras obtained here admit adjustments similar to that of the string \(L_\infty\)-algebra \cite{Samann:2019eei}, which is a central extension of a simple Lie algebra.

\paragraph{Organisation of this paper.}
This paper is organised as follows. After a brief summary of the language of \(L_\infty\)-algebras in \cref{eq:linfty-review}, we review kinematical Lie algebras and classify their iterated \(L_\infty\)-algebraic central extensions in \cref{sec:classification}.
Then we discuss the gravitational theories obtained by gauging such kinematical \(L_\infty\)-algebras in \cref{sec:gravity}, and explain how branes may couple to them in \cref{sec:brane}.

\paragraph{Notational conventions.}
We use the Koszul sign rule throughout. The notation \(V[i]\) denotes suspension of a \(\mathbb Z\)-graded vector space \(V\) such that \(V[i]^j = V^{i+j}\). The notation \(\bigodot V\) denotes the graded-symmetric algebra generated by \(V\).

\section{Lightning review of \texorpdfstring{\(L_\infty\)-}{𝐿∞‐}algebras}\label{eq:linfty-review}
Let us briefly review \(L_\infty\)-algebras (sometimes called `(strongly) homotopy Lie algebras') to establish terminology and conventions. More detailed reviews may be found in \cite{Jurco:2018sby,Kraft:2022efy,Lada:2021vvm,Lada:1992wc}.

An \(L_\infty\)-algebra is a \(\mathbb Z\)-graded real vector space
\begin{equation}
    \mathfrak g=\bigoplus_{i\in\mathbb Z}\mathfrak g^i
\end{equation}
equipped with totally graded-antisymmetric \(i\)-ary multilinear maps
\begin{equation}
    \mu_i\colon\overbrace{\mathfrak g\otimes\dotsb\otimes\mathfrak g}^i\to\mathfrak g
\end{equation}
of degree \(2-i\) obeying the following homotopy Jacobi identity:
\begin{equation}\label{eq:homotopy_jacobi_identity}
    \sum_{\substack{i+j=k\\\sigma\in\operatorname{Sym}(k)}}
    \frac{(-1)^{ij}}{i!j!}\chi(\sigma)
    \mu_{i+1}(\mu_i(x_{\sigma(1)},\dotsc,x_{\sigma(i)}),x_{\sigma(i+1)},\dotsc,x_{\sigma(k)})
    =0
\end{equation}
for all integers \(k\ge1\), where \(\sigma\) ranges over permutations of \(\{1,\dotsc,k\}\) and where \(\chi(\sigma)\) is the graded-antisymmetric Koszul sign defined such that
\begin{equation}
    x_{\sigma(1)}\wedge\dotsb\wedge x_{\sigma(k)}\eqqcolon\chi(\sigma)x_1\wedge\dotsb\wedge x_k
\end{equation}
with \(\wedge\) being graded-antisymmetric. An \(L_\infty\)-algebra in which \(\mu_i=0\) except for \(\mu_2\) is the same as a Lie algebra, and the homotopy Jacobi identity \eqref{eq:homotopy_jacobi_identity} reduces to the ordinary Jacobi identity for \(\mu_2\). An \(L_\infty\)-algebra in which \(\mu_i=0\) except for \(\mu_1\) and \(\mu_2\) is the same as a differential graded Lie algebra, where \(\mu_1\) is the differential and \(\mu_2\) is the Lie bracket.

When \(\mathfrak g\) is finite-dimensional (an assumption that holds for all \(L_\infty\)-algebras in this paper), the data of an \(L_\infty\)-algebra can equivalently be encoded in the \emph{Chevalley--Eilenberg algebra} \(\operatorname{CE}(\mathfrak g)\), which is the free \(\mathbb Z\)-graded unital graded-commutative associative algebra
\begin{equation}
    \operatorname{CE}(\mathfrak g) = \bigodot\mathfrak g^*[-1]
\end{equation}
together with a differential \(\mathrm d\) defined on generators \(t^a\in\mathfrak g^*[-1]\) as
\begin{equation}
    \mathrm dt^a = \sum_{i=1}^\infty \frac1{i!}f^a_{b_1\dotso b_i}t^{b_1}\dotso t^{b_i},
\end{equation}
where \(f^a_{b_1\dotso b_i}\) are the structure constants of \(\mu_i\),
and extended to the rest of \(\operatorname{CE}(\mathfrak g)\) via the graded Leibniz rule. A \emph{morphism} \(\phi\colon\mathfrak f\to\mathfrak g\) between \(L_\infty\)-algebras is then given by a morphism of unital differential graded algebras \(\operatorname{CE}(\mathfrak g)\to\operatorname{CE}(\mathfrak f)\) in the opposite direction.

An element \(x\in\mathfrak g\) of an \(L_\infty\)-algebra is \emph{central} if all brackets involving it vanish:
\begin{equation}
    \mu_i(x,y_1,\dotsc,y_{i-1})=0
\end{equation}
for any \(i\in\mathbb Z^+\) and \(y_1,\dotsc,y_{i-1}\in\mathfrak g\). A \emph{central extension} of an \(L_\infty\)-algebra \(\mathfrak g\) is a short exact sequence
\begin{equation}
    0\to V\to \tilde{\mathfrak g}\to\mathfrak g\to0
\end{equation}
where \(V\) is a vector space regarded as an \(L_\infty\)-algebra with all brackets vanishing and such that the image of \(V\) in \(\tilde{\mathfrak g}\) is central.

The \emph{cohomology} of an \(L_\infty\)-algebra \(\mathfrak g\) is the cohomology of its Chevalley--Eilenberg algebra \(\operatorname{CE}(\mathfrak g)\); it agrees with Lie algebra cohomology when \(\mathfrak g\) is concentrated in degree zero.

The cohomology of an \(L_\infty\)-algebra classifies its central extensions; in particular, if the \(L_\infty\)-algebra is in fact a Lie algebra \(\mathfrak g\) (concentrated in degree zero), a \(k\)th cohomology class \([x_1\dotsm x_k]\) with \(x_1,\dotsc,x_k\in\mathfrak g\) corresponds to a central element in degree \(2-k\) expressible as \(\mu_k(x_1,\dotsc,x_k)\). This generalises the usual statement that the second Lie algebra cohomology \(\operatorname H^2(\mathfrak g)\) classifies its Lie-algebra central extensions.

\section{Classification of rotation-independent \texorpdfstring{\(L_\infty\)-}{𝐿∞‐}algebraic central extensions of kinematical Lie algebras}\label{sec:classification}
We first determine which \((d+1)\)-dimensional non-Lorentzian spacetime algebras admit appropriate \(L_\infty\)-algebraic central extensions. The class of non-Lorentzian spacetime algebras we example is called \emph{kinematical Lie algebras} \cite[Def.~1]{Bergshoeff:2022eog}, which are Lie algebras of spatial rotations \(\mathfrak o(d)\) (with generators \(\mathtt j_{ij}\)), boosts (with generators \(\mathtt t^1_i\)), spatial translations (with generators \(\mathtt t^2_i\)), and time translation (with generator \(\mathtt h\)) such that all generators transform as expected \(\mathfrak o(d)\) representations; this class includes Poincaré, (anti-)de~Sitter, Galilean, and Carrollian groups along with others, but does not include (for example) symmetry algebras that break \(\mathfrak o(d)\) spatial rotation symmetry.

We are concerned with \(L_\infty\)-algebraic central extensions whose corresponding cocycles do \emph{not} involve the spatial rotation generators \(\mathtt j_{ij}\). This is the natural non-Lorentzian analogue of the corresponding restriction in the super-Poincaré brane bouquet \cite{Fiorenza:2013nha} where one only deals with cocycles that do not depend on spacetime rotation generators \(\mathtt j_{\mu\nu}\).

\subsection{Review of the classification of kinematical Lie algebras}
A kinematical Lie algebra is a Lie algebra on the same underlying vector space as that of the Poincaré or Galilean algebras, but in which only rotation is guaranteed to work `correctly'. They have been classified in arbitrary numbers of spacetime dimensions, falling into several infinite families that exist in arbitrary spacetime dimension \(d+1\) and a handful of exceptional ones in \(d\le3\). We review their definition and classification below.

\begin{definition}[{\cite[Def.~1]{Bergshoeff:2022eog}}]\label{def:kinematical_lie_algebra}
    A \emph{kinematical Lie algebra} in \(d+1\) spacetime dimensions is a Lie algebra \(\mathfrak g\) whose underlying vector space is
    \begin{equation}
        \mathfrak g = \mathfrak o(d)\times(\mathbb R^2\otimes\mathbb R^d)\times\mathbb R
    \end{equation}
    with basis elements \(\mathtt j_{ij},\mathtt t^a_i,\mathtt h\) (with \(i,j\in\{1,\dotsc,d\}\), \(a\in\{1,2\}\), and \(\mathtt j_{ij}=-\mathtt j_{ji}\))
    such that \(\mathfrak o(d)\) is a Lie subalgebra and such that the brackets between \(\mathfrak o(d)\) and 
    \(\mathbb R^d\oplus\mathbb R^d\oplus\mathbb R\) are
    \begin{align}\label{eq:kinematic_lie_algebra_brackets}
        [\mathtt j_{ij},\mathtt t^a_k]&=\delta_{jk}\mathtt t^a_i-\delta_{ik}\mathtt t^a_j,&
        [\mathtt j_{ij},\mathtt h]&=0.
    \end{align}
    (That is, \(\mathtt t^1\) and \(\mathtt t^2\) are vectors while \(\mathtt h\) is a scalar.)
\end{definition}
In the case of the Poincaré or Galilean algebras, \(\mathtt j_{ij}\) corresponds to spatial rotations, \(\mathtt t^1_i\) and \(\mathtt t^2_i\) to boosts and spatial translations, and \(\mathtt h\) to time translations.
This definition does not constrain the brackets between \(\mathtt t\) and \(\mathtt h\) except through the Jacobi identity. The kinematical Lie algebras in \(d\ge4\) (or, equivalently, those infinite families of kinematical Lie algebras that exist for every \(d\)) are classified as follows \cite{Figueroa-OFarrill:2017tcy,Figueroa-OFarrill:2017sfs}.
\begin{proposition}[\cite{Figueroa-OFarrill:2017tcy,Figueroa-OFarrill:2017sfs}]\label{prop:kinematic_lie_algebra_classification}
In \(d\ge4\), a kinematical Lie algebra is isomorphic to one of the following Lie algebras:
\begin{itemize}
\item a simple Lie algebra, more specifically one of the anti-de Sitter group \(\mathfrak o(d+1,1)\), de Sitter group \(\mathfrak o(d,2)\), or the Euclidean orthogonal group \(\mathfrak o(d+2)\);
\item the Poincaré algebra \(\mathfrak o(d,1)\ltimes\mathbb R^d\) or the Euclidean Poincaré algebra \(\mathfrak o(d+1)\ltimes\mathbb R^d\);
\item the \emph{Carrollian algebra} \(\mathfrak{carr}(d) \coloneqq \mathfrak o(d)\ltimes\mathfrak{heis}(d)\), where \(\mathfrak{heis}(d)\) is the \((2d+1)\)-dimensional Heisenberg Lie algebra with
\begin{equation}
    [\mathtt t^a_i,\mathtt t^b_j]=\delta_{ij}\varepsilon^{ab}\mathtt h
\end{equation}
as its only nonzero Lie bracket;
\item or a \emph{generalised Newton--Hooke algebra} \(\mathfrak{newt}(d;M)\coloneqq\mathfrak o(d)\ltimes(\mathbb R\ltimes(\mathbb R^2\otimes\mathbb R^d))\) where \(M\) is a \(2\times2\) real matrix and \(\mathbb R^2\otimes\mathbb R^d\) (with generators \(\mathtt t^a_i\)) is an Abelian Lie algebra upon which the Abelian Lie algebra \(\mathbb R\) (with generator \(\mathtt h\)) acts as
\begin{equation}
    [\mathtt h,\mathtt t^a_i]=M^a{}_b\mathtt t^b_i.
\end{equation}
Furthermore \(M\) can always be put in a canonical form as either \((\begin{smallmatrix}0&0\\0&0\end{smallmatrix})\) (static algebra), \((\begin{smallmatrix}0&1\\0&0\end{smallmatrix})\) (Galilean algebra),
\((\begin{smallmatrix}1&1\\0&1\end{smallmatrix})\), \((\begin{smallmatrix}\gamma&0\\0&1\end{smallmatrix})\) for \(-1\le\gamma\le1\), or \((\begin{smallmatrix}\chi&1\\-1&\chi\end{smallmatrix})\) for \(\chi\ge0\).
\end{itemize}
\end{proposition}
The above list of kinematical Lie algebras are all obtainable as Wigner--İnönü deformations of a simple Lie algebra except for the generalised Newton--Hooke algebras \(\mathfrak{newt}(d;M)\) with \(\operatorname{tr}M\ne0\).

There are exactly four generalised Newton--Hooke algebras that are obtainable as such Wigner--İnönü deformations: the static algebra \(\mathfrak{newt}(d;(\begin{smallmatrix}0&0\\0&0\end{smallmatrix}))\), the Galilean algebra \(\mathfrak{gal}(d)\coloneqq\mathfrak{newt}(d;(\begin{smallmatrix}0&1\\0&0\end{smallmatrix}))\), and the two Newton--Hooke algebras \(\mathfrak{newt}(d;(\begin{smallmatrix}-1&0\\0&1\end{smallmatrix}))\) and \(\mathfrak{newt}(d;(\begin{smallmatrix}0&1\\-1&0\end{smallmatrix}))\). These are also precisely those that admit a nontrivial Lie-algebraic central extension; the central extension is then \([\mathtt t_i^a,\mathtt t_j^b]=\delta_{ij}\varepsilon^{ab}\mathtt z\).

In \(d\le3\), there exist additional kinematical Lie algebras beyond the ones that exist in every dimension as given in \cref{prop:kinematic_lie_algebra_classification}; let us call them \emph{exceptional} kinematical Lie algebras. They are classified as follows.
\begin{proposition}[\cite{Figueroa-OFarrill:2017sfs,Figueroa-OFarrill:2017ycu}]\label{prop:exceptional_kinematic_3d}
    In \(d=3\), a kinematical Lie algebra is either isomorphic to one of the Lie algebras given in \cref{prop:kinematic_lie_algebra_classification} or to one of the following:
    \begin{itemize}
        \item the Lie algebra \(\mathbb3_1^\pm\coloneqq\mathbb R\oplus\mathfrak k^\pm\), where \(\mathfrak k_\pm\) has the same underlying vector space as \(\mathfrak o(3)\ltimes(\mathfrak o(3)\oplus\mathbb R^3)\) (spanned by \(\mathtt j_{ij},\mathtt t^1_i,\mathtt t^2_i\)) but differs from it solely by the additional Lie bracket
    \begin{equation}
        [\mathtt t^2_i,\mathtt t^2_j] = \pm(\varepsilon_{ijk}\mathtt t^1_k-\mathtt j_{ij})
    \end{equation}
    \cite[(50,51)]{Figueroa-OFarrill:2017ycu}.
    The Lie algebra \(\mathfrak k_-\) is a semidirect product \((\mathfrak o(3)\otimes\mathbb R[s]/(s^2-1))\ltimes\mathfrak o(3)\), where the left factor \(\mathfrak o(3)\otimes\mathbb R[s]/(s^2-1)\) is spanned by \(\mathtt j_{ij}\) and \(\mathtt t^1_i+\mathtt t^2_i\) (with the former spanning \(\mathfrak o(3)\) and the latter spanning \(s\mathfrak o(3)\)), and the right factor \(\mathfrak o(3)\) is spanned by \(\mathtt t^1_i\), with the action given by the adjoint action of \(\mathfrak o(3)\) composed with the quotient \(\mathfrak o(3)\otimes\mathbb R[s]/(s^2)\twoheadrightarrow\mathfrak o(3)\) induced by the ring homomorphism \(\operatorname{ev}_1\colon \mathbb R[s]/(s^2-1)\twoheadrightarrow\mathbb R,\;s\mapsto1\). The Lie algebra \(\mathfrak k_+\) is a real form of the complexification of \(\mathfrak k_-\).
        \item \(\mathbb 3_1^0\coloneqq\mathfrak o(3)\ltimes(\mathfrak o(3)\oplus\mathbb R^3\oplus\mathbb R)\), where \(\mathfrak o(3)\oplus\mathbb R^3\oplus\mathbb R\) corresponds to \(\mathtt t^1_i\) and \(\mathtt t^2_i\) and \(\mathtt h\) respectively \cite[(52)]{Figueroa-OFarrill:2017ycu}
        \item \(\mathbb 3_2\coloneqq\mathfrak o(3)\ltimes(\widetilde{\mathbb R^3}\oplus\mathbb R)\) \cite[(53)]{Figueroa-OFarrill:2017ycu}
        \item \(\mathbb 3_3\coloneqq\mathfrak o(3)\ltimes(\mathfrak o(3)\oplus (\mathbb R^3\rtimes\mathbb R))\), where the Abelian Lie algebra \(\mathbb R^3\) is acted upon by the Abelian Lie algebra \(\mathbb R\) with unit weight \cite[(62)]{Figueroa-OFarrill:2017ycu}
        \item \(\mathbb 3_4\coloneqq\mathfrak o(3)\ltimes(\widetilde{\mathbb R^3}\rtimes\mathbb R)\), where \(\mathbb R\) acts on \(\widetilde{\mathbb R^3}\) as \([\mathtt h,\mathtt t^1_i]=-\mathtt t^2_i\) \cite[(83)]{Figueroa-OFarrill:2017ycu}
        \item \(\mathbb 3_5\coloneqq\mathfrak o(3)\ltimes(\widetilde{\mathbb R^3}\rtimes\mathbb R)\), where \(\mathbb R\) acts on \(\widetilde{\mathbb R^3}\) as \([\mathtt h,\mathtt t^1_i]=\mathtt t^1_i\) and \([\mathtt h,\mathtt t^2_i]=2\mathtt t^2_i\) \cite[(64)]{Figueroa-OFarrill:2017ycu}
    \end{itemize}
    In the above, \(\widetilde{\mathbb R^3}\) is the Lie algebra universal central extension \(\operatorname H^2(\mathbb R^3)\to\widetilde{\mathbb R^3}\to\mathbb R^3\) of the Abelian Lie algebra \(\mathbb R^3\), that is, \(\widetilde{\mathbb R^3}\) is spanned by \(\mathtt t^a_i\) with \([\mathtt t^1_i,\mathtt t^1_j]=\varepsilon_{ijk}\mathtt t^2_k\) and no other nonzero Lie brackets.
\end{proposition}
In \cite[Table~4]{Figueroa-OFarrill:2017sfs} and \cite[Table~1]{Figueroa-OFarrill:2017ycu}, they are listed (below a line) in the order \(\mathbb3_1^+,\mathbb3_1^-,\mathbb3_1^0,\mathbb3_2,\mathbb3_3,\mathbb3_4,\mathbb3_5\).

\begin{proposition}[\cite{Figueroa-OFarrill:2017sfs,Andrzejewski:2018gmz}]
    In \(d=2\), a kinematical Lie algebra is either isomorphic to one of the Lie algebras given in \cref{prop:kinematic_lie_algebra_classification} or to one of the following:
    \begin{itemize}
        \item \(\mathfrak{newt}(2;M,\tilde M)\coloneqq(\mathfrak o(2)\oplus\mathbb R)\ltimes(\mathbb R^2\otimes\mathbb R^2)\) where the Abelian Lie algebra \(\mathbb R^2\otimes\mathbb R^2\) is acted upon by \(\mathbb R\) as
        \begin{equation}
            [\mathtt h,\mathtt t^a_i]
            =M^a{}_b\mathtt t^b_i
            +
            \tilde M^a{}_b\varepsilon_{ij}\mathtt t^b_j
        \end{equation}
        for some \(2\times2\) matrices \(M\) and \(\tilde M\) \cite[(6), (48), (52), (56)]{Andrzejewski:2018gmz}. Furthermore, \(M\) and \(\tilde M\) can always be taken to be \(M=(\begin{smallmatrix}1&0\\0&\lambda\end{smallmatrix})\) and \(\tilde M=(\begin{smallmatrix}0&0\\0&\theta\end{smallmatrix})\) with \(-1<\lambda\le1\) and \(\theta\in\mathbb R\).
        \item the Lie algebra \(\mathbb2_1\coloneqq\mathfrak o(2)\ltimes \mathfrak{heis}(2)\), where the \(\mathfrak{heis}(2)\) is realised as \([\mathtt t^a_i,\mathtt t^b_j]=\delta^{ab}\varepsilon_{ij}\mathtt h\) \cite[(79)]{Andrzejewski:2018gmz}.
        \item the Lie algebra \(\mathbb2_2\coloneqq\mathfrak o(3)\ltimes \mathfrak o(3)\), where the left \(\mathfrak o(3)\) is spanned by
        \(\mathtt j_{12},\mathtt t^1_1,\mathtt t^1_2\) and the right \(\mathfrak o(3)\) is spanned by \(\mathtt h,\mathtt t^2_1,\mathtt t^2_2\) and the action is the adjoint action \cite[(83)]{Andrzejewski:2018gmz}.\footnote{
            The translation to the notation used in \cite[(83)]{Andrzejewski:2018gmz} is
            \(\mathtt j_{12}=R\), \((\mathtt t^1_1,\mathtt t^1_2)=(\Re(P-B),\Im(P-B))\),
            \(\mathtt h=H\), and
            \((\mathtt t^2_1,\mathtt t^2_2)=(\Re B,\Im B)\).
        }
        \item the Lie algebras \(\mathfrak o(2)\ltimes(\mathfrak h\oplus\mathbb R^2)\) where \(\mathfrak h\) is one of \(\mathfrak o(3)\) or \(\mathfrak{sl}(2)\) or \(\mathfrak{heis}(1)\) (we write \(\mathbb2_3^+\coloneqq\mathfrak o(2)\ltimes(\mathfrak o(3)\oplus\mathbb R^2)\), \(\mathbb2_3^+\coloneqq\mathfrak o(2)\ltimes(\mathfrak{sl}(2)\oplus\mathbb R^2)\), \(\mathbb2_3^0\coloneqq\mathfrak o(2)\ltimes(\mathfrak{heis}(1)\oplus\mathbb R^2)\)), which is spanned by \(\mathtt t^1_i\) and \(\mathtt h\) and realised as
        \begin{align}
            [\mathtt t^1_i,\mathtt t^1_j]&=\varepsilon_{ij}\mathtt h,&
            [\mathtt h,\mathtt t^1_i]=\sigma\varepsilon_{ij}\mathtt t^1_j
        \end{align}
        with \(\sigma\in\{+1,-1,0\}\).
        (The choice \(+1\) leads to \(\mathfrak o(3)\) 
         \cite[(68)]{Andrzejewski:2018gmz}; the choice \(-1\) leads to \(\mathfrak{sl}(2;\mathbb R)\) 
         \cite[(68)]{Andrzejewski:2018gmz}; the choice \(0\) leads to \(\mathfrak{heis}(1)\) \cite[(63)]{Andrzejewski:2018gmz}.)
        \item the Lie algebra \(\mathbb2_4\coloneqq\mathfrak o(2)\ltimes\widetilde{\mathfrak{heis}}(1)\) where \(\widetilde{\mathfrak{heis}}(1)\) is the universal central extension \(\mathbb R^2\cong\operatorname H^2(\mathfrak{heis}(1))\to \widetilde{\mathfrak{heis}}(1)\to\mathfrak{heis}(1)\) of the Heisenberg algebra \(\mathfrak{heis}(1)\), given by the Lie brackets
        \begin{align}
            [\mathtt t^1_i,\mathtt t^1_j]&=\varepsilon_{ij}\mathtt h,&
            [\mathtt h,\mathtt t^1_i]=\mathtt t^2_i
        \end{align}
        and all others vanishing \cite[(66)]{Andrzejewski:2018gmz}.
    \end{itemize}
\end{proposition}
In \cite[Table~3]{Figueroa-OFarrill:2017sfs}, they are listed (below a line) in the order \(\mathbb2_1,\mathbb2_2,\mathbb2_3^0,\mathbb2_4,\mathbb2_3^\pm\) except for \(\mathfrak{newt}(2;(\begin{smallmatrix}1&0\\0&\lambda\end{smallmatrix}),(\begin{smallmatrix}0&0\\0&\theta\end{smallmatrix}))\), which is listed above the dividing line;  
\cite[Table~1]{Andrzejewski:2018gmz} follows the same order except that 
\(\mathfrak{newt}(2;(\begin{smallmatrix}1&0\\0&\lambda\end{smallmatrix}),(\begin{smallmatrix}0&0\\0&\theta\lambda\end{smallmatrix}))\) is listed below the dividing line.

For completeness, we mention that in \(d=1\) any three-dimensional Lie algebra is a \(d=1\) kinematical Lie algebra, so the classification reduces to the Bianchi classification \cite{1898MMFSI..11..267B,Bianchi:2001pja} of three-dimensional Lie algebras; in \(d=0\), the only kinematical Lie algebra is the Abelian one-dimensional Lie algebra \(\mathbb R\).

\subsection{\texorpdfstring{\(L_\infty\)-}{𝐿∞‐}algebra central extensions of kinematical Lie algebras for \texorpdfstring{\(d\ge4\)}{𝑑≤4}}
\(L_\infty\)-algebra extensions are computed by Lie algebra cohomology, i.e.\ the cohomology of the Chevalley--Eilenberg algebra \(\operatorname{CE}(\mathfrak g)\cong\bigodot\mathfrak g^*[-1]\). We assume that we are working with a non-exceptional kinematical Lie algebra (i.e.\ belonging to the families that can be defined for any \(d\)).
By abuse of notation we use the same symbols for the basis of \(\mathfrak g\) and the basis of \(\mathfrak g^*\) except that the indices are located in the opposite locations; thus \(\mathfrak g^*\) is spanned by \(\mathtt h\), \(\mathtt t^i_a\), and \(\mathtt j^{ij}\). We also assume \(d\ge2\).

Given a cochain \(X^{i_1i_2\dotso}\) built out of \(\mathtt t^i_a\) and \(\mathtt h\), its coboundary always contains terms corresponding to its \(\mathfrak o(d)\) transformation:
\[
    \mathrm dX^{i_1i_2\dotso} \sim \mathtt j^{i_1j}X^{ji_2\dotso} + \mathtt j^{i_2j}X^{i_1j\dotso} + \dotsb.
\]
Thus, a necessary condition for the cochain \(X^{i_1i_2\dotso}\) to be a cocycle is that it must be invariant under \(\mathfrak o(d)\).
This proves to be a boon, for (assuming \(d\ge2\)) there exist only a small set of \(\mathfrak o(d)\)-invariant combinations of \(\mathtt h\), \(\mathtt t^i_a\), and \(\mathtt j^{ij}\). Namely, any such \(\mathfrak o(d)\)-invariant \(X\) is a polynomial of the following \(\mathfrak o(d)\)-invariant cochains:
\begin{align}
    \mathtt h&,&
    \mathtt x
    &\coloneqq\mathtt t^i_a\mathtt t^j_b\delta_{ij}\varepsilon^{ab},&
    y_{a_1\dotso a_d}
    =
    y_{(a_1\dotso a_d)}
    \coloneqq
    \frac1{d!}\varepsilon^{i_1\dotso i_d}\mathtt t_{a_1}^{i_1}\dotsm\mathtt t_{a_d}^{i_d}.
\end{align}
(Due to grading, other terms such as \(\mathtt t^1_i\mathtt t^1_j\delta^{ij}=0\) vanish.) For brevity, let us introduce the index \(A=(a_1a_2\dotso a_d)\) to write \(y_A\); then \(A\in\{0,1,\dotsc,d\}\) is the index for the \((d+1)\)-dimensional representation of the group \(\operatorname{SL}(2;\mathbb R)\) that rotates between \(\mathtt t^1_i\) and \(\mathtt t^2_i\). Note that \(\mathtt h^2=x^{d+1}=xy^A=0\) and that \(y_Ay_B\) vanishes except for a component proportional to \(x^d\). Hence the space of all polynomials of \(\mathtt h,x,y_A\) is spanned by the basis
\begin{equation}
    \left\{x^k,\mathtt hx^k,y_A,\mathtt hy_A\middle|k\in\{0,1\dotsc,d\},\;A\in\{0,\dotsc,d\}\right\}.
\end{equation}

From the algebra, in the case \(d\ge4\) (or for any \(d\ge2\) as long as one does not consider one of the exceptional kinematical Lie algebras in \(d\in\{2,3\}\)), the differentials may be parameterised as
\begin{align}
    \mathrm d\mathtt h &= \alpha x,&
    \mathrm dx&= \beta\mathtt hx,&
    \mathrm dy_A &= \mathtt hy_BN^B{}_A\end{align}
for some numbers \(\alpha\) and \(\beta\) and a matrix \(N^B{}_A\). Now,
\begin{align}
    \mathrm d^2\mathtt h &= \alpha\beta\mathtt hx,
    &
    \mathrm d^2x&=\alpha\beta x^2,&
    \mathrm d^2y^A&=0.
\end{align}
The Chevalley--Eilenberg cochain complex is then
\begin{equation}
\begin{tikzcd}[cramped, column sep=small]
    \mathtt h \rar["\alpha"] &x \rar["\beta"] & \mathtt hx \rar["\alpha"] & x^2 \rar["2\beta"] &{}& \hskip-2.6em\dotsb \rar ["(d-1)\beta"] & \mathtt hx^{d-1}\rar["\alpha"] & x^d \rar["d\beta"] & \mathtt hx^d \\
    &&&& y_A \ar[r, "N^B{}_A"] &\mathtt hy_B.
\end{tikzcd}
\end{equation}
Hence nilpotence of \(\mathrm d\) requires \(\alpha\beta=0\), and we have the three cases \(\alpha\ne0=\beta\), \(\alpha=0\ne\beta\), and \(\alpha=0=\beta\), for each of which it is easy to work out the cohomology.
\begin{itemize}
\item When \(\alpha\ne0=\beta\) and \(N\) is nondegenerate (as for the simple, Poincaré or Euclidean cases), then the only coclosed cocycle that is not obviously coexact is \(\mathtt hx^d\). In these cases, however, it can be seen by inspection that it is in fact coexact if one also considers cochains containing \(\mathtt j_{ij}\).
\item When \(\alpha\ne0=\beta\) and \(N=0\) (this is the Carrollian algebra \(\mathfrak{carr}(d)\), then the \(\mathtt j\)-free cohomology is given by the cocycles \(\mathtt hx^d\) (in degree \(2d+1\)) and \(y_A\) (in degree \(d\)) and \(\mathtt hy_A\) (in degree \(d+1\)).
\begin{equation}
\begin{tikzcd}[cramped, column sep=small]
    \mathtt h \rar["\alpha"] &x & \mathtt hx \rar["\alpha"] & x^2 && \hskip-2.6em\dotsb & \mathtt hx^{d-1}\rar["\alpha"] & x^d & \mathtt hx^d \\
    &&&& y_A &\mathtt hy_B.
\end{tikzcd}
\end{equation}
\item When \(\alpha=0\ne\beta\) (as for the generalised Newton--Hooke algebra \(\mathfrak{newt}(d;M)\) with \(\operatorname{tr}M\ne0\)), then the \(\mathtt j\)-free cohomology is given by the 1-cocycle \(\mathtt h\) in addition to
\(\ker N\) (in degree \(d\), corresponding to the \(d\)-cocycle \(c^Ay_A\) with \(c\in\ker N\)) and \(\operatorname{coker}N\) (in degree \(d+1\), corresponding to the \((d+1)\)-cocycle \(\tilde c^A\mathtt hy_A\) for arbitrary \(\tilde c\) modulo those where \(\tilde c\in\operatorname{im}N\)). (The cocycles \(\mathtt hx^k\propto\mathrm d(x^k)\) are coexact for \(k>0\).)
\begin{equation}
\begin{tikzcd}[cramped, column sep=small]
    \mathtt h &x \rar["\beta"] & \mathtt hx & x^2 \rar["2\beta"] &{}& \hskip-2.6em\dotsb \rar ["(d-1)\beta"] & \mathtt hx^{d-1} & x^d \rar["d\beta"] & \mathtt hx^d \\
    &&&& y_A \ar[r, "N^B{}_A"] &\mathtt hy_B.
\end{tikzcd}
\end{equation}
\item When \(\alpha=0=\beta\) (as for the generalised Newton--Hooke algebra \(\mathfrak{newt}(d;M)\) with \(\operatorname{tr}M=0\), including the Galilean algebra \(\mathfrak{newt}(d;(\begin{smallmatrix}0&1\\0&0\end{smallmatrix}))\)), then the \(\mathtt j\)-free cohomology is given by \(\{x,x^2,\dotsc,x^d,\mathtt h,\mathtt hx,\dotsc,\mathtt hx^d\}\) in addition to the \(d\)-cocycle \(c_Ay^A\) with \(c\in\ker N\) and  the \((d+1)\)-cocycle \(\tilde c_A\mathtt hy^A\) with \(\tilde c\in\operatorname{coker}N\).
\begin{equation}
\begin{tikzcd}[cramped, column sep=small]
    \mathtt h &x  & \mathtt hx & x^2  &{}& \hskip-2.6em\dotsb & \mathtt hx^{d-1} & x^d & \mathtt hx^d \\
    &&&& y_A \ar[r, "N^B{}_A"] &\mathtt hy_B.
\end{tikzcd}
\end{equation}
\end{itemize}

Thus we have shown the following proposition.
\begin{proposition}\label{prop:jfree_cohomology_nonexceptional}
    The \(\mathtt j_{ij}\)-free cohomology of a kinematical Lie algebra \(\mathfrak g\) in \(d\ge4\) is as follows.
    \begin{itemize}
        \item If \(\mathfrak g\) is simple, Poincaré or Euclidean, there is no \(\mathtt j\)-free cohomology.
        \item If \(\mathfrak g=\mathfrak{carr}(d)\) is Carrollian, then the cocycles are \(\mathtt hx^d\) and \(y^{(a_1\dotso a_d)}\) and \(\mathtt hy^{(a_1\dotso a_d)}\), so that the maximal \(\mathtt j_{ij}\)-free \(L_\infty\)-algebraic central extension has underlying graded vector space
        \begin{equation}
            \tilde{\mathfrak g}=\mathfrak g\times\mathbb R^{d+1}[d-2]\times\mathbb R^{d+1}[d-1]\times\mathbb R[2d-1]
        \end{equation}
        with extra central generators \(\mathtt z_{(2-d)}^{a_1\dotso a_d}\) (in degree \(2-d\)) and \(\mathtt z_{(1-d)}^{a_1\dotso a_d}\) (in degree \(1-d\) and \(\mathtt a_{(1-2d)}\) (in degree \(1-2d\)) such that
        \begin{equation}
        \begin{aligned}
            \mu_d(\mathtt t^{a_1}_{i_1},\dotsc,\mathtt t^{a_d}_{i_d})&=\varepsilon_{i_1\dotso i_d}\mathtt z_{(2-d)}^{a_1\dotso a_d},\\
            \mu_{d+1}(\mathtt h,\mathtt t^{a_1}_{i_1},\dotsc,\mathtt t^{a_d}_{i_d})&=\varepsilon_{i_1\dotso i_d}\mathtt z_{(1-d)}^{a_1\dotso a_d},\\
            \mu_{2d+1}(\mathtt h,\mathtt t^1_1,\dotsc,\mathtt t^1_d,\mathtt t^2_1,\dotsc,\mathtt t^2_d)&=\mathtt a_{(1-2d)}.
        \end{aligned}
        \end{equation}
        \item If \(\mathfrak g=\mathfrak{newt}(d;M)\) is a generalised Newton--Hooke algebra and \(\operatorname{tr}M\ne 0\), then the cocycles are as above but without \(\mathtt h\) and \(x\), so that
        \begin{equation}
            \tilde{\mathfrak g}
            =\mathfrak g\times(\ker N)[d-2]\times(\operatorname{coker}N)[d-1]
        \end{equation}
        with extra central generators \(\mathtt z^\alpha_{(2-d)}\) (in degree \(2-d\), valued in \(\ker N\)), and \(\mathtt z_{(1-d)}^\beta\) (in degree \(1-d\), valued in \(\operatorname{coker}N\)) such that 
        \begin{equation}
        \begin{aligned}
            \mu_d(\mathtt t^{a_1}_{i_1},\dotsc,\mathtt t^{a_d}_{i_d})&=\varepsilon_{i_1\dotso i_d}P^{a_1\dotso a_d}_\alpha \mathtt z_{(2-d)}^\alpha,\\
            \mu_{d+1}(\mathtt h,\mathtt t^{a_1}_{i_1},\dotsc,\mathtt t^{a_d}_{i_d})&=\varepsilon_{i_1\dotso i_d}\tilde P^{a_1\dotso a_d}_\alpha \mathtt z_{(1-d)}^\beta,
        \end{aligned}
        \end{equation}
        where \(P\) and \(\tilde P\) are projectors to \(\ker N\) and \(\operatorname{coker}N\) respectively.
        \item If \(\mathfrak g=\mathfrak{newt}(d;M)\) is a generalised Newton--Hooke algebra and \(\operatorname{tr}M=0\), then the cocycles are the cocycles are \(\mathtt h\), \(x\) and
        \(c_{a_1\dotso a_d}y^{(a_1\dotso a_d)}\) for
        \(\ker N\) and
        \(\tilde c_{a_1\dotso a_d}\mathtt hy^{(a_1\dotso a_d)}\) for
        \(\tilde c\in\operatorname{coker}N\). (That is, the cocycles \(\tilde c_{a_1\dotso a_d}\mathtt hy^{(a_1\dotso a_d)}\) for which \(\tilde c\in\operatorname{im}N\) are coboundaries.) The maximal \(\mathtt j\)-free \(L_\infty\)-algebraic central extension has underlying graded vector space
        \begin{equation}
            \tilde{\mathfrak g}
            =\mathfrak g\times\left(
            \bigoplus_{i=-1}^{2d-1}
                \mathbb R[i]
            \right)\times(\ker N)[d-2]\times(\operatorname{coker}N)[d-1]
        \end{equation}
        with extra central generators \(\mathtt a_{(1)},\dotsc,\mathtt a_{(1-2d)}\) (in degrees \(1,0,\dotsc,1-2d\) respectively), \(\mathtt z_{(2-d)}^\alpha\) (in degree \(2-d\) taking values in \(\ker N\)), and \(\mathtt z_{(1-d)}^\beta\) (in degree \(1-d\) taking values in \(\operatorname{coker}N\)) such that 
        \begin{equation}
        \begin{aligned}
            \mu_{2k}(\mathtt t_{\hat\imath_1},\mathtt t_{\hat\jmath_1},\dotsc,\mathtt t_{\hat\imath_k},\mathtt t_{\hat\jmath_k})&=
            \delta_{[\hat\imath_1\hat\jmath_1}\dotsm
            \delta_{\hat\imath_k\hat\jmath_k]}\mathtt a_{(2-2k)},\\
            \mu_{2k+1}(\mathtt h,\mathtt t_{\hat\imath_1},\mathtt t_{\hat\jmath_1},\dotsc,\mathtt t_{\hat\imath_k},\mathtt t_{\hat\jmath_k})&=
            \delta_{[\hat\imath_1\hat\jmath_1}\dotsm
            \delta_{\hat\imath_k\hat\jmath_k]}\mathtt a_{(1-2k)},\\
            \mu_d(\mathtt t^{a_1}_{i_1},\dotsc,\mathtt t^{a_d}_{i_d})&=\varepsilon_{i_1\dotso i_d}P^{a_1\dotso a_d}_\alpha \mathtt z_{(2-d)}^\alpha,\\
            \mu_{d+1}(\mathtt h,\mathtt t^{a_1}_{i_1},\dotsc,\mathtt t^{a_d}_{i_d})&=\varepsilon_{i_1\dotso i_d}\tilde P^{a_1\dotso a_d}_\alpha \mathtt z_{(1-d)}^\beta,
        \end{aligned}
        \end{equation}
        where \(\hat\imath,\hat\jmath,\dotsc\in\{1,\dotsc,2d\}\) are indices combining \((a,i)\in\{1,2\}\times\{1,\dotsc,d\}\), and \(\delta_{\hat\imath\hat\jmath}\coloneqq\varepsilon^{ab}\delta_{ij}\) where \(\hat\imath=(a,i)\) and \(\hat\jmath=(b,j)\),
        and
        where \(P\) and \(\tilde P\) are projectors to \(\ker N\) and \(\operatorname{coker}N\) respectively.
    \end{itemize}
\end{proposition}

\subsection{\texorpdfstring{\(L_\infty\)-}{𝐿∞‐}algebra central extensions of kinematical Lie algebras for \texorpdfstring{\(d\in\{2,3\}\)}{𝑑∈\{2, 3\}}}
For the exceptional kinematical Lie algebras in \(d\in\{2,3\}\), we compute the \(\mathtt j\)-free cohomology by direct calculation.
In what follows, various nonzero constants have been omitted where they do not affect the computation of the cohomology.

For \(d=3\), the possible \(\mathfrak o(3)\)-invariant \(\mathtt j\)-free cochains are
\begin{equation}
\begin{tikzcd}[column sep=tiny, row sep=tiny]
\mathtt h  & x & \mathtt hx \\
&& y^A & \mathtt hy^B  \\
&&& x^2 & \mathtt hx^2 & x^3 & \mathtt hx^3.
\end{tikzcd}
\end{equation}
The differentials amongst them for the \(d=3\) exceptional kinematical Lie algebras are then as follows.
\begin{itemize}
\item \(\mathbb3_1^\pm\): \(\mathrm d\mathtt h=0\), \(\mathrm dx=y^{(1)}\pm y^{(3)}\), \(\mathrm dy^{(0)}=\pm x^2\), \(\mathrm dy^{(2)}=x^2\), \(\mathrm dy^{(1)}=\mathrm dy^{(3)}=0\).
\begin{equation}\label{eq:3_1_diagram}
\begin{tikzcd}
\mathtt h  & x \drar["\delta^1_A\pm\delta^3_A"'] & \mathtt hx \drar["-(\delta^1_A\pm\delta^3_A)"] \\
&& y^A \drar["\delta^A_2\pm\delta^A_0"'] & \mathtt hy^B \drar["-(\delta^B_2\pm\delta^B_0)"] \\
&&& x^2 & \mathtt hx^2 & x^3 & \mathtt hx^3 \\
\end{tikzcd}
\end{equation}
In this case, using \(\mathrm d\mathtt j_{ij}=\mp\mathtt t^2_i\mathtt t^2_j+\dotsb\), the fact that \(\mathrm d(\varepsilon^{ijk}\mathtt j_{ij}\mathtt t^2_k)\propto y^{(3)}\) and \(\mathrm d(\mathtt h\varepsilon^{ijk}\mathtt j_{ij}\mathtt t^2_k)\propto\mathtt hy^{(3)}\) kills some of the would-be cohomology components in the diagram \eqref{eq:3_1_diagram}.
\item \(\mathbb3_1^0\): \(\mathrm d\mathtt h=0\), \(\mathrm dx=y^{(1)}\), \(\mathrm dy^A=0\) except for \(\mathrm dy^{(2)}=x^2\).
\begin{equation}
\begin{tikzcd}
\mathtt h  & x  \drar["\delta^1_A"'] & \mathtt hx \drar["-\delta^1_B"] \\
&& y^A \drar["\delta^A_2"'] & \mathtt hy^B \drar["-\delta^B_2"] \\
&&& x^2  & \mathtt hx^2 & x^3  & \mathtt hx^3 \\
\end{tikzcd}
\end{equation}
\item \(\mathbb3_2\): \(\mathrm d\mathtt h=0\), \(\mathrm dx=y^{(0)}\), \(\mathrm dy^A=0\) except for \(\mathrm dy^{(3)}=x^2\).
\begin{equation}
\begin{tikzcd}
\mathtt h  & x \drar["\delta^0_A"'] & \mathtt hx \drar["-\delta^0_B"] \\
&& y^A \drar["\delta^A_3"'] & \mathtt hy^B \drar["-\delta^A_3"] \\
&&& x^2 & \mathtt hx^2 & x^3  & \mathtt hx^3 \\
\end{tikzcd}
\end{equation}
\item \(\mathbb3_3\): \(\mathrm d\mathtt h=0\), \(\mathrm dx=x+y^{(1)}\), \(\mathrm dy^A=A\mathrm dy^A+\delta^A_2x^2\).
\begin{equation}
\begin{tikzcd}
\mathtt h  & x \rar["1"] \drar["\delta^1_A"'] & \mathtt hx \drar["-\delta^1_B"] \\
&& y^A \drar["\delta^A_2"'] \rar["A\delta^B_A"] & \mathtt hy^B \drar["-\delta^B_2"] \\
&&& x^2 \rar["2"'] & \mathtt hx^2 & x^3 \rar["3"'] & \mathtt hx^3 \\
\end{tikzcd}
\end{equation}
\item \(\mathbb3_4\): \(\mathrm d\mathtt h=0\), \(\mathrm dx=y^{(0)}\), \(\mathrm dy^A=A\mathrm dy^{A-1}+\delta^A_3x^2\).
\begin{equation}
\begin{tikzcd}
\mathtt h  & x \drar["\delta^0_A"'] & \mathtt hx \drar["-\delta^0_B"] \\
&& y^A \drar["\delta^A_3"'] \rar["A\delta^A_{B+1}"] & \mathtt hy^B \drar["-\delta^B_3"] \\
&&& x^2 & \mathtt hx^2 & x^3 & \mathtt hx^3 \\
\end{tikzcd}
\end{equation}
\item \(\mathbb3_5\): \(\mathrm d\mathtt h=0\), \(\mathrm dx=\mathtt hx+y^{(0)}\), \(\mathrm dy^A=N^A{}_B\mathtt hy^A_B+\delta^A_3x^2\), where the matrix \(N^A{}_B\) is diagonal and of full rank.
\begin{equation}
\begin{tikzcd}
\mathtt h  & x \rar["1"] \drar["\delta^0_A"'] & \mathtt hx \drar["-\delta^0_B"] \\
&& y^A \drar["\delta^A_3"'] \rar["N^A{}_B"] & \mathtt hy^B \drar["-\delta^B_3"] \\
&&& x^2 \rar["2"'] & \mathtt hx^2 & x^3 \rar["3"'] & \mathtt hx^3 \\
\end{tikzcd}
\end{equation}
\end{itemize}
For \(d=2\), the possible \(\mathfrak o(2)\)-invariant \(\mathtt j\)-free cochains are
\begin{equation}
\begin{tikzcd}[column sep=tiny, row sep=tiny]
\mathtt h &y^A  & \mathtt hx \\
& x & \mathtt hy^B & x^2 & \mathtt hx^2.
\end{tikzcd}
\end{equation}
The differentials amongst them for the \(d=2\) exceptional kinematical Lie algebras are then as follows.
\begin{itemize}
\item \(\mathbb2_1\): here \(\mathrm d\mathtt h=y^{(0)}+y^{(2)}\) while \(\mathrm dx=0\) and \(\mathrm dy^A=0\).
\begin{equation}
\begin{tikzcd}
\mathtt h \ar[r, "\delta^0_A+\delta^2_A"] &y^A  & \mathtt hx \\
& x & \mathtt hy^B \rar["\delta^B_0+\delta^B_2"'] & x^2 & \mathtt hx^2
\end{tikzcd}
\end{equation}
\item \(\mathbb2_2\): \(\mathrm d\mathtt h=y^{(0)}+y^{(2)}\), and \(\mathrm dx=\mathtt hy^{(1)}\), and \(\mathrm dy^{(0)}=\mathrm dy^{(2)}=0\) and \(\mathrm dy^{(1)}=\mathtt hx\).
\begin{equation}
\begin{tikzcd}
\mathtt h \ar[r, "\delta^0_A+\delta^2_A"] &y^A \rar["\delta^1_A"] & \mathtt hx \\
& x\rar["\delta^1_B"'] & \mathtt hy^B \rar["\delta^B_0+\delta^B_2"']& x^2 & \mathtt hx^2
\end{tikzcd}
\end{equation}
\item \(\mathbb2_3^0\):  \(\mathrm d\mathtt h=y^{(0)}\) and \(\mathrm dx=0=\mathrm dy^A\).
\begin{equation}
\begin{tikzcd}
\mathtt h \ar[r, "\delta^0_A"] &y^A & \mathtt hx \\
& x & \mathtt hy^B \rar["\delta^B_2"'] & x^2 & \mathtt hx^2
\end{tikzcd}
\end{equation}
\item \(\mathbb2_3^\pm\): \(\mathrm d\mathtt h=y^{(0)}\) and \(\mathrm dx=\pm\mathtt h y^{(1)}\) and \(\mathrm dy^{(0)}=\mathrm dy^{(2)}=0\) and \(\mathrm dy^{(1)}=\pm\mathtt hx\).
\begin{equation}
\begin{tikzcd}
\mathtt h \ar[r, "\delta^0_A"] &y^A \rar["\pm\delta^1_A"] & \mathtt hx \\
& x\rar["\pm\delta^1_B"'] & \mathtt hy^B \rar["\delta^B_2"'] & x^2 & \mathtt hx^2\\
\end{tikzcd}
\end{equation}
\item \(\mathbb2_4\): \(\mathrm d\mathtt h=y^{(0)}\) and \(\mathrm dx=0\) and \(\mathrm dy^{(0)}=0\) and \(\mathrm dy^{(1)}=\mathtt hy^{(0)}\) and \(\mathrm dy^{(2)}=\mathtt hy^{(1)}\).
\begin{equation}
\begin{tikzcd}
\mathtt h \ar[r, "\delta^0_A"] &y^A \drar["\delta^A_{B+1}"] & \mathtt hx \\
& x & \mathtt hy^B \rar["\delta^B_2"'] & x^2  & \mathtt hx^2\\
\end{tikzcd}
\end{equation}
\item \(\mathfrak{newt}(2;M,\tilde M)\):
we have \(\mathrm d\mathtt h=0\) and, since
\begin{equation}
    \mathrm dt^a_i=M^a{}_bt^b_i+\varepsilon_{ij}\tilde M^a{}_bt^b_j,
\end{equation}
then
\begin{equation}
\begin{aligned}
    \mathrm dx &= 2(M^a{}_bt^b_i+\varepsilon_{ij}\tilde M^a{}_bt^b_j)\varepsilon_{ac}t^c_i
    = 2 \delta^{ij} (\varepsilon_{ab}M^c{}_b)t^b_i t^a_j
    + 2 \varepsilon_{ij}\varepsilon_{ac}\tilde M^a{}_bt^b_jt^c_i\\
    &= 2 (\operatorname{tr}M) \mathtt hx + 
    (\varepsilon\tilde M)_{(ab)}\mathtt hy^{ab}
\end{aligned}
\end{equation}
and
\begin{equation}
    \mathrm dy^{ab} \sim \tilde M^{(a|}{}_c\varepsilon^{c|b)}x +
    M^a{}_c \mathtt hy^{cb}.
\end{equation}
Hence
\begin{equation}
\begin{tikzcd}
\mathtt h &y^A \drar["M"', pos=0] \rar["\varepsilon\tilde M"] & \mathtt hx \\
& x\rar["\varepsilon\tilde M"']\urar["\operatorname{tr}M"', very near end] & \mathtt hy^B & x^2 \rar & \mathtt hx^2.
\end{tikzcd}
\end{equation}
The cohomologies in degrees \(2\) and \(3\) are then given in terms of the kernel and cokernel respectively of the \(4\times4\) matrix
\begin{equation}\label{eq:hatM-definition}
    \hat M \coloneqq
    \begin{pmatrix}
        M & \varepsilon \tilde M \\
        \varepsilon\tilde M & 2\operatorname{tr}M
    \end{pmatrix},
\end{equation}
where indices have been omitted.
\end{itemize}

\begin{landscape}
\begin{table}
\begin{center}
\begin{tabular}{cccccccc}\toprule
\(\mathfrak g\)
&\(\dim\operatorname H^1_\text{\(\mathtt j\)-free}(\mathfrak g)\)
&\(\dim\operatorname H^2_\text{\(\mathtt j\)-free}(\mathfrak g)\)
&\(\dim\operatorname H^3_\text{\(\mathtt j\)-free}(\mathfrak g)\)
&\(\dim\operatorname H^4_\text{\(\mathtt j\)-free}(\mathfrak g)\)
&\(\dim\operatorname H^5_\text{\(\mathtt j\)-free}(\mathfrak g)\)
&\(\dim\operatorname H^6_\text{\(\mathtt j\)-free}(\mathfrak g)\)
&\(\dim\operatorname H^7_\text{\(\mathtt j\)-free}(\mathfrak g)\) \\\midrule
\(\mathbb3_1^\pm\) & 1 & 0 & 1 & 1 & 0 & 1 & 1\\
\(\mathbb3_1^0\) & 1 & 2 & 2 & 2 & 0 & 1 & 1 \\
\(\mathbb3_2\) & 1 & 0 & 2 & 2 & 0 & 1 & 1\\
\(\mathbb3_3\) & 1 & 0 & 1 & 1 & 0 & 0 & 0\\
\(\mathbb3_4\) & 1 & 0 & 2 & 2 & 0 & 1 & 1\\
\(\mathbb3_5\) & 1 & 0 & 0 & 0 & 0 & 0 & 0\\\midrule
\(\mathbb2_1\) & 0 & 3 & 3 & 0 & 1 \\
\(\mathbb2_2\) & 0 & 0 & 0 & 0 & 1\\
\(\mathbb2_3^0\) & 0 & 3 & 3 & 0 & 1 \\
\(\mathbb2_3^\pm\) & 0 & 1 & 0 & 0 & 1\\
\(\mathbb2_4\) & 0 & 1 & 1 & 0 & 1\\
\(\mathfrak{newt}(2;M,\tilde M)\) & 1 & \(\ker\hat M\) & \(\operatorname{coker}\hat M\) & 0 & 0 \\
\bottomrule
\end{tabular}
\end{center}
\caption{The \(\mathtt j\)-free cohomologies of the exceptional kinematical Lie algebras in \(d\in\{2,3\}\).
The matrix \(\hat M\) is defined in \eqref{eq:hatM-definition}.
}\label{table:exceptional_cohomology}
\end{table}
\end{landscape}

Thus, by direct computation we have shown the following.
\begin{proposition}\label{prop:jfree_cohomology_exceptional}
The \(\mathtt j\)-free cohomology of the exceptional kinematical Lie algebras in \(d=2\) and \(d=3\) are as given in \cref{table:exceptional_cohomology}.
\end{proposition}

\subsection{Iterated central extensions: the non\texorpdfstring-{‐}Lorentzian brane bouquet}
The preceding sections classified \(\mathtt j\)-free central extensions of the kinematical Lie algebras. However, iterated central extensions (i.e.\ \(L_\infty\)-algebras obtained by repeatedly taking central extensions) arise in a number of contexts such as the brane bouquet (see \cref{ssec:bouquet_review}). In this section we classify all \(\mathtt j\)-free iterated central extensions of kinematical Lie algebras subject to the assumption that each central extension is of degree \(<d\). This additional assumption is computationally convenient but also corresponds to the physical assumption that all branes involved must have codimension greater than two (in a \(d+1\)-dimensional spacetime). Codimension-two branes are subject to braiding statistics and, due to logarithmic divergences, have complicated backreaction \cite{deBoer:2012ma,Greene:1989ya}.

According to \cref{prop:jfree_cohomology_nonexceptional}, for any non-exceptional kinematical Lie algebra \(\mathfrak g\), there are no nontrivial \(\mathtt j\)-free cocycles of degree \(<d\) except for \(\mathfrak g=\mathfrak{newt}(d;M)\). Hence there are three remaining cases: exceptional kinematical Lie algebras in \(d\in\{2,3\}\), generalised Newton--Hooke algebras \(\mathfrak{newt}(d;M)\) with \(\operatorname{tr}M\ne0\), and generalised Newton--Hooke algebras \(\mathfrak{newt}(d;M)\) with \(\operatorname{tr}M=0\). We consider each of the three cases in turn.

\paragraph{Exceptional kinematical Lie algebras.} In the exceptional cases, \cref{prop:jfree_cohomology_exceptional} shows that the only \(\mathtt j\)-free cocycle of degrees \(<d\) is \(\mathtt h\) for all three-dimensional exceptional kinematical Lie algebras (\(\mathbb3_1^\pm,\mathbb3_1^0,\mathbb3_2,\mathbb3_3,\mathbb3_4,\mathbb3_5\)) as well as \(\mathfrak{newt}(2;M,\tilde M)\).
In these cases, we introduce a new central generator \(\mathtt a\) of degree \(1\) such that
\begin{align}
    \mathrm d\mathtt a&=\mathtt h,&
    \mathrm d\mathtt h&=0.
\end{align}
Then, for any univariate polynomial \(p(-)\), the expression \(p(\mathtt a)\mathtt h\) is a new cocycle. However, it is coexact since \(p(\mathtt a)\mathtt h=\mathrm d(q(\mathtt a))\) where \(q\) is any antiderivative of \(p\).

\paragraph{Generalised Newton--Hooke (nonzero trace).}
Similarly, for \(\mathfrak{newt}(d;M)\) with \(\operatorname{tr}M\ne0\), the only nontrivial \(\mathtt j\)-free cocycle of degree \(<d\) is again \(\mathtt h\), so that one can introduce a new central generator \(\mathtt a\) with
\begin{align}
    \mathrm d\mathtt a&=\mathtt h,&
    \mathrm d\mathtt h&=0.
\end{align}
as before. Using it, one obtains cocycles \(p(\mathtt a)\mathtt hx^k\) with \(k\ge0\). However, if \(k>0\), since
\begin{equation}
    p(\mathtt a)\mathtt hx^k
    =
    k^{-1}\beta^{-1}\mathrm d(p(\mathtt a)x^k)
    -
    k^{-1}\beta^{-1}p'(\mathtt a)\mathtt hx^k
\end{equation}
(where \(\mathrm dx=\beta\mathtt hx\)), we see that up to coboundaries and nonzero overall constants
\begin{equation}
    p(\mathtt a)\mathtt hx^k \sim p'(\mathtt a)\mathtt hx^k
    \sim p''(\mathtt a)\mathtt hx^k\sim\dotsb\sim 0,
\end{equation}
so that \(p(\mathtt a)\mathtt hx^k\) is in fact coexact. Even the \(k=0\) case is coexact since \(p(\mathtt a)\mathtt h=\mathrm d(q(\mathtt a))\) where \(q\) is any antiderivative of \(p\).

\paragraph{Generalised Newton--Hooke (traceless).}
For \(\mathfrak{newt}(d;M)\) with \(\operatorname{tr}M\ne0\), the nontrivial \(\mathtt j\)-free cocycle of degree \(<d\) are of the form \(x^k\) or \(\mathtt hx^k\). For both of these types one can introduce new central generators
\begin{align}
    \mathrm d\mathtt a_{(1-2k)}&=\mathtt hx^k,&
    \mathrm d\mathtt a_{(2-2k)}&=x^k.
\end{align}
Now, if \(p\) is a polynomial consisting solely of \(\mathtt a_{(1)},\mathtt a_{(-1)},\mathtt a_{(-3)},\dotsc\) as well as \(x\), then
\begin{equation}
    p(\mathtt a_{(1)},\mathtt a_{(-1)},\dotsc,x)\mathtt h
\end{equation}
is a cocycle; this is no longer the case if \(p\) also depends on \(\mathtt a_{(0)},\mathtt a_{(-2)},\dotsc\). So then we may introduce, for a polynomial \(p\), another central extension with a central generator
\begin{equation}
    \mathrm d\mathtt b_p = p(\mathtt a_{(1)},\mathtt a_{(-1)},\dotsc,x)\mathtt h
\end{equation}
and so on. That is, (ignoring \(\mathtt a_{(1)},\mathtt a_{(-1)},\mathtt a_{(-3)},\dotsc\) entirely) the space of \(k\)th iterated central extensions seems to be putatively given by the vector space \(V_k\) defined iteratively as
\begin{align}
    V_0 &= \mathbb R[-2],&
    V_1 &= \bigodot V_0,&
    V_2 &= \bigodot (V_1\oplus V_0),&
    \!\!\!\dotsc,
    V_i &= \bigodot (V_{i-1}\oplus\dotsb\oplus V_0).
\end{align}
(We can ignore the complication that \(x^{d+1}=0\) since we are only concerned with cocycles of degree \(<d\).)
Note that there are two canonical families of maps, which are embeddings except for \(\tilde\iota_{0\to j}\):
\begin{equation}
    \iota_{i\to j},\tilde\iota_{i\to j}\colon V_i \to V_j\qquad(i<j),
\end{equation}
where \(\iota_{i\to j}\) takes values in monomials
\begin{equation}
    V_i\hookrightarrow \bigodot V_i \hookrightarrow \bigodot (V_{j-1}\oplus\dotsb\oplus \underbrace{V_i}\oplus\dotsb\oplus V_0) = V_j,
\end{equation}
whereas \(\tilde\iota_{i\to j}\) instead maps \(V_i\) to polynomials:
\begin{equation}
    V_i=\bigodot(V_{i-1}\oplus \dotsb\oplus V_0)\hookrightarrow \bigodot (V_{j-1}\oplus\dotsb\oplus \underbrace{V_{i-1}\oplus\dotsb\oplus V_0}) = V_j\qquad(i>0),
\end{equation}
and we define \(\tilde\iota_{0\to j}=0\) for later convenience.

The space \(V_k\) is too large, however, since we have not quotiented out by those cocycles that are coexact. The coboundary operator maps a generator \(\iota_{i\to j}(v)\) (with \(v\in V_i\)) to its derivative, which is nothing other than the polynomial \(\tilde\iota_{i\to j}(v)\). That is, for \(j>0\), define
\begin{equation}
    \deltait_j\colon V_j\to V_j
\end{equation}
as the derivation defined on the generators of \(V_j=\mathbb R[V_{j-1},\dotsc,V_0]\) as
\begin{equation}
    \deltait_i\colon \iota_{i\to j}(v)\mapsto \tilde\iota_{i\to j}(v) \qquad(v\in V_i).
\end{equation}
(We set \(\delta_0\colon V_0\to V_0\) to be identically zero for later convenience.) Then we should be quotienting out \(\delta_j(V_j)\) from \(V_j\) since these are coexact cocycles.

Moreover, this means that cocycles that depend on would-be generators that should not exist (because they correspond to cocycles that are not in fact coexact) should be quotiented out as well. That is, we must quotient out also by the ideal
\begin{equation}
    (\iota_{i-1\to i}(\deltait_{i-1}(V_{i-1}))+\dotsb+\iota_{0\to i}(\deltait_0(V_0)))V_i
    \subset V_i
\end{equation}
since there are in fact no generators corresponding to \(\deltait_{i-1}(V_{i-1}),\dotsc,\deltait_1(V_1)\). (The term \(\iota_{0\to i}(\deltait_0(V_0))\) is harmless since it is identically zero.) Thus, the correct space of iterated central extensions is given by
\begin{equation}\label{eq:tildeV_definition}
    \tilde V_i = V_i / (\deltait_i(V_i) + (\iota_{i-1\to i}(\deltait_{i-1}(V_{i-1}))+\dotsb+\iota_{0\to i}(\deltait_0(V_0)))V_i),
\end{equation}
where the quotient and the sum are those of graded vector spaces, not rings.

Thus we have shown the following.
\begin{proposition}\label{prop:iterated_classification}
For any kinematical Lie algebra except for \(\mathfrak{newt}(d;M)\) with \(\operatorname{tr}M=0\), there are no nontrivial \(\mathtt j\)-free, degree \(<d\) iterated central extensions. For \(\mathfrak{newt}(d;M)\) with \(\operatorname{tr}M=0\), the space of \(\mathtt j\)-free, degree \(<d\) \(k\)th iterated central extensions for \(k>1\) is in canonical bijection with the degree \(<d\) components of \(\tilde V_k\) as defined in \eqref{eq:tildeV_definition}.
\end{proposition}

\section{Non\texorpdfstring-{‐}Lorentzian gravities from \texorpdfstring{\(L_\infty\)-}{𝐿∞‐}algebra extensions of kinematical Lie algebras}\label{sec:gravity}
It is well known that one can construct the kinematic data for Einstein gravity and Newtonian gravity by starting with gauging the Poincaré or Bargmann algebras and imposing certain constraints on the curvatures.
In this section, we apply this procedure to the \(L_\infty\)-algebraic central extensions of kinematical Lie algebras to obtain kinematic data for gravitational theories. The fact that the algebras involved are more general \(L_\infty\)-algebras than Lie algebras results in the fact that we obtain differential forms of various degrees in addition to the usual one-forms (vielbein, spin connection, etc.).

We restrict ourselves to constructing the kinematical data; we do not write down action principles or equations of motion. (For non-Lorentzian gravitational theories, there may be obstructions to writing down action principles \cite{Bergshoeff:2015uaa}.)

\subsection{Review of Poincaré and Bargmann gravities}
There is a uniform procedure to construct the kinematic data of gravitational theories associated to a kinematical Lie algebra \(\mathfrak g\): namely, one considers a principal \(G\)-bundle\footnote{Often one takes this bundle to be trivial so that the resulting kinematic data are equivalent to a second-order metric-based formalism, but summing over principal bundles may be sometimes advantageous, cf.\ the discussion in \cite{Borsten:2025phf}.} (where \(G\) exponentiates \(\mathfrak g\)) with a connection \(A\in\Omega^1(M;\mathfrak g)\), and constrains some components of the field strength \(F\in\Omega^2(M;\mathfrak g)\) to vanish.

For the Poincaré case \(\mathfrak g=\mathfrak{io}(d,1)\), we have the connection
\begin{equation}
    (e^a,\omega^{ab})\in\Omega^1(M;\mathfrak g)
\end{equation}
where \(a,b,\dotsc\) are \(d+1\)-dimensional internal Lorentz indices. The field strength
\begin{equation}
    \operatorname{Curv} = (T^a,R^{ab})
\end{equation}
consists of the torsion \(T\) and the Riemann tensor \(R\). We constrain \(T=0\) (a so-called `conventional' constraint since derivatives of \(\omega\) do not appear in \(T\)); then for generic values of \(A\), the torsion-freeness condition \(T=0\) implies that the spin connection \(\omega\) may be solved in terms of the vielbein \(e\) so that the only independent field left is the vielbein on \(M\) as expected.

For the Bargmann algebra \(\mathfrak g=\mathfrak{barg}(d)\) \cite{Andringa:2010it,Andringa:2013mma}, the Lie-algebraic central extension of the Galilean algebra \(\operatorname{newt}(d;(\begin{smallmatrix}0&1\\0&0\end{smallmatrix}))\), we again have the potential
\begin{equation}
    (\omega^{ij},e^i,\varpi^i,\tau,A^{(1)})
\end{equation}
corresponding to \((\mathtt j^{ij},\mathtt t^i_1,\mathtt t^i_2,\mathtt h,\mathtt a_{(0)})\) where \(\mathtt a_{(0)}\) is the central extension.
We impose the constraints that the curvature components for \(\mathtt h\), \(\mathtt t^i_1\), and \(\mathtt a_{(0)}\) vanish. This lets us solve for \(\omega^{ij}\) and \(\varpi^i\) in terms of \(\tau\), \(e^a\), and \(A^{(1)}\).
In particular, the curvatures are
\begin{equation}
\begin{aligned}
    \operatorname{Curv}[\mathtt h] &= \mathrm d\tau \\
    \operatorname{Curv}[\mathtt t^i_1] &= \mathrm de^i + \omega^{ij}\wedge e_j + \varpi^i\wedge\tau, \\
    \operatorname{Curv}[\mathtt j^{ij}] &= \mathrm d\omega^{ij} + \omega^{ik}\wedge\omega_k{}^j, \\
    \operatorname{Curv}[\mathtt t^i_2] &= \mathrm d\varpi^i + \omega^i{}_j\wedge\varpi^j, \\
    \operatorname{Curv}[\mathtt a_{(0)}] &= \mathrm dA^{(1)} + \varpi^i\wedge e^j\delta_{ij}.
\end{aligned}
\end{equation}
If we constrain \(\operatorname{Curv}[\mathtt h]=0\), this means that \(\tau\) is locally exact, so that we can locally write
\begin{equation}\label{eq:localtime}
    \tau = \mathrm dt.
\end{equation}
Then \(t\) defines local absolute time, foliating spacetime \(M\) into spatial slices \(M_t\), and \(e^i_\mu e_{i\nu}\) defines the spatial metric on the slices \(M_t\).

If we further impose \(\operatorname{Curv}[J^{ij}]=0\), then the Riemannian metric on the slices \(M_t\) is flat. Then one can gauge-fix all components of \(e^a,\tau,A^{(1)}\) except for one component of \(A^{(1)}\); examining how the one-form couples to the particle worldline then shows that this is indeed the Newtonian gravitational potential.

\subsection{The two extremes: Newton--Hooke versus Carrollian}
According to \eqref{prop:jfree_cohomology_nonexceptional}, there are two kinds of cocycles that can appear in the non-exceptional kinematical Lie algebras:
\begin{itemize}
\item For the generalised Newton--Hooke algebras \(\mathfrak{newt}(d;M)\) with \(\operatorname{tr}M=0\) (namely, the static algebra, the Galilean algebra, and the two Newton--Hooke algebras), there are cocycles \(\mathtt a_{(i)}\) of degrees \(1,2,\dotsc,2d+1\) that correspond to differential-form potentials of form degrees \(0,\dotsc,2d\). Of course, on a \((d+1)\)-dimensional spacetime, differential forms of degree greater than \(d+1\) vanish. (For the Carrollian algebra there exists a cocycle of degree \(2d+1\) of this form, corresponding to a \(2d\)-form potential, but this vanishes for degree reasons.)
\item For the Carrollian algebra \(\mathfrak{car}(d)\) and for the generalised Newton--Hooke algebra \(\mathfrak{newt}(d;M)\) where \(M\) is degenerate, there are cocycles \(\mathtt z_{(2-d)}\) and \(\mathtt z_{(1-d)}\), of degrees \(d\) and \(d+1\) respectively, corresponding to \((d-1)\)-form and \(d\)-form potentials respectively.
\end{itemize}
Thus, the two Newton--Hooke algebras and the Carrollian algebra exhibit different features: the former has a series of \(p\)-forms of all degrees \(p\) that generalised the one-form potential of Newton--Cartan gravity; the latter has \((d-1)\) and \(d\)-forms of a different kind altogether. The Galilean and static cases combine features of both of these extremes.
In what follows, therefore, we present the two extreme cases separately: the former in \cref{ssec:case1}, the latter in \cref{ssec:carroll_gravity}.

\subsubsection{The universal sector of maximally extended generalised Newton--Hooke gravity with \texorpdfstring{\(\operatorname{tr}M=0\)}{tr 𝑀=0}}\label{ssec:case1}
We define the kinematic data
\begin{equation}
\begin{aligned}
    \operatorname{Curv}[\mathtt j^{ij}]
    &=\mathrm d\omega^{ij}+\omega^{ik}\wedge\omega_k{}^j,\\
    \operatorname{Curv}[\mathtt t^i_a]
    &=\mathrm de^i_a+\omega^i{}_j\wedge e_a^j+M^b{}_a\tau e_b^i,\\
    \operatorname{Curv}[\mathtt h]
    &=\mathrm d\tau,\\
    \operatorname{Curv}[\mathtt z_{(1)}]
    &=\mathrm dA^{(0)}+\tau,\\
    \operatorname{Curv}[\mathtt z_{(0)}]
    &=\mathrm dA^{(1)}+e^i_a\wedge e^j_b\delta_{ij}\varepsilon^{ab},\\
    \operatorname{Curv}[\mathtt z_{(-1)}]
    &=\mathrm dA^{(2)}+\tau\wedge e^i_a\wedge e^j_b\delta_{ij}\varepsilon^{ab},\\    
    \operatorname{Curv}[\mathtt z_{(-2)}]
    &=\mathrm dA^{(3)}+e^i_a\wedge e^j_b\wedge
    e^k_c\wedge e^l_d\delta_{ij}\delta_{kl}\varepsilon^{ab}\varepsilon^{cd},\\
    \vdots
\end{aligned}
\end{equation}
One can constrain some or all of these curvatures (at least, those that are \emph{conventional}, i.e.\ don't involve derivatives of the spin connection) to be zero or to some fixed torsion, similar to what is done for torsionful Newton--Cartan gravity \cite{Geracie:2015dea,Bergshoeff:2017dqq,Figueroa-OFarrill:2020gpr} that arises in Lifshitz holography \cite{Christensen:2013lma,Christensen:2013rfa,Bergshoeff:2014uea}, in certain limits of Einstein gravity \cite{VandenBleeken:2017rij}, and the quantum Hall effect \cite{Geracie:2014nka}.

In particular, suppose that one constrains \(\operatorname{Curv}[\mathtt z_{(1)}]\) to be zero. This forces \(\tau\) to be exact, with an antiderivative given by \(-A^{(0)}\), which then defines a global time function, similar to \eqref{eq:localtime}; whereas in Newton--Cartan gravity the closedness of \(\tau\) only lets one define a \emph{local} time function \(t\) (absent assumptions about topology, i.e.\ the vanishing of the first de~Rham cohomology of spacetime), here \(A^{(0)}\) defines a true global time function.

Similarly, constraining \(\operatorname{Curv}[\mathtt z_{(1-p)}]\) means that the corresponding \(p\)-form potential \(A^{(p)}\) is fixed completely in terms of \((e^i_1,\tau)\) (the usual spatiotemporal vielbein) as well as \(e^i_2\). Concretely, let us impose
\begin{equation}
    0=\operatorname{Curv}[\mathtt z_{(1-p)}].
\end{equation}
These are all conventional constraints in the sense that they do not depend on derivatives of the spin connection \(\omega^{ij}\) or \(e^1_i\) (or, for that matter, the vielbein \(e^2_j\) or \(\tau\)). If \(p=1+2k\) is odd, then this means
\begin{equation}
    0=\mathrm dA^{(1+2k)}-\overbrace{\mathrm dA^{(1)}\wedge\dotsm\wedge\mathrm dA^{(1)}}^{k+1},
\end{equation}
which can be solved as
\begin{equation}\label{eq:gauge-odd}
    A^{(1+2k)}=A^{(1)}\wedge\overbrace{\mathrm dA^{(1)}\wedge\dotsm\wedge\mathrm dA^{(1)}}^k.
\end{equation}
If \(p=2k+2\) is even, then this means
\begin{equation}
    0=\mathrm dA^{(2k)}-\mathrm dA^{(0)}\wedge\overbrace{\mathrm dA^{(1)}\wedge\dotsm\wedge\mathrm dA^{(1)}}^{k+1},
\end{equation}
which can be solved as
\begin{equation}
    A^{(2k)}=A^{(0)}\wedge\overbrace{\mathrm dA^{(1)}\wedge\dotsm\wedge\mathrm dA^{(1)}}^k
\end{equation}
or (if one prefers a gauge that does not diverge as the time coordinate \(A^{(0)}\) tends to infinity) as
\begin{equation}\label{eq:gauge-even}
    A^{(2k)}=-\mathrm dA^{(0)}\wedge A^{(1)}\wedge\overbrace{\mathrm dA^{(1)}\wedge\dotsm\wedge\mathrm dA^{(1)}}^{k-1}.
\end{equation}
Thus, the \(A^{(p)}\) for \(p\ge2\) are then all fixed in terms of \(A^{(1)}\) (the Newton--Cartan gravitational potential) and \(A^{(0)}\) (the global time function) up to gauge choices, and there are no further geometric structures beyond the gravitational potential \(A^{(1)}\) the same way that the spin connection is determined by the vielbein in Einstein gravity.

\subsubsection{Maximally extended Carrollian gravity}\label{ssec:carroll_gravity}
In the Carrollian case, representing in a sense the opposite extreme to the Newton--Hooke case, we have (in addition to the spatial vielbein \(e^i_1\), spacetime spin connection \(e^i_2\), spatial spin connection \(\omega^{ij}\) and temporal vielbein \(\tau\), which are all one-forms), according to \cref{prop:jfree_cohomology_nonexceptional}, a \(d\)-cocycle \(\mathtt z_{(2-d)}^{a_1\dotso a_d}\) (with dual basis element \(\mathtt z^{(2-d)}_{a_1\dotso a_d}\)) corresponding to a \((d-1)\)-form potential \(A^{(d-1)}\),
a \((d+1)\)-cocycle  \(\mathtt z_{(1-d)}^{a_1\dotso a_d}\) (corresponding to the dual basis element \(\mathtt z^{(1-d)}_{a_1\dotso a_d}\)) corresponding to a \(d\)-form potential \(A^{(d)}\),
and a \((2d+1)\)-cocycle \(\mathtt a_{(1-2d)}\) that would correspond to a \(2d\)-form potential, which vanishes for degree reasons on a \((d+1)\)-dimensional spacetime.
From the brackets of the \(L_\infty\)-algebra, we may then read off the corresponding curvatures as follows,
\begin{equation}
\begin{aligned}
    \operatorname{Curv}[\mathtt j^{ij}]
    &=\mathrm d\omega^{ij}+\omega^{ik}\wedge\omega_k{}^j,\\
    \operatorname{Curv}[\mathtt t^i_a]
    &=\mathrm de^i_a+\omega^{ij}\wedge e^k_a\delta_{jk},\\
    \operatorname{Curv}[\mathtt h]
    &=\mathrm d\tau+e^i_a\wedge e^j_b\delta_{ij}\varepsilon^{ab},\\
    \operatorname{Curv}[\mathtt z^{(2-d)}_{a_1\dotso a_d}]
    &=
    \mathrm dA^{(d-1)}+e^{i_1}_{a_1}\wedge\dotsb\wedge e^{i_d}_{a_d}\varepsilon_{i_1\dotso i_d},\\
    \operatorname{Curv}[\mathtt z^{(1-d)}_{a_1\dotso a_d}]
    &=
    \mathrm dA^{(d)}+\tau\wedge e^{i_1}_{a_1}\wedge\dotsb\wedge e^{i_d}_{a_d}\varepsilon_{i_1\dotso i_d}.
\end{aligned}
\end{equation}
We see that the curvatures \(\operatorname{Curv}[\mathtt z^{(2-d)}_{a_1\dotso a_d}]\) and \(\operatorname{Curv}[\mathtt z^{(1-d)}_{a_1\dotso a_d}]\) function as `covariantised' versions of the spatial volume \(d\)-form and the spatiotemporal volume \((d+1)\)-form respectively. Alternatively, if one constrains these curvatures to zero, then \(A^{(d-1)}\) and \(A^{(d)}\) correspond to potentials for the spatial or spacetime volume forms respectively.

\section{Brane actions and couplings to gravity}\label{sec:brane}
Given the gravitational field content found in \cref{sec:gravity}, the natural follow-up question is to determine how these fields couple to matter.
Given the presence of \(p\)-form fields and the brane bouquet, it is natural to consider couplings to branes, which we explore in this section.

\subsection{Lightning review of brane actions and the brane bouquet}\label{ssec:bouquet_review}
The \(p\)-form fields of various supergravity theories may be described by central extensions \(\mathfrak g\) of the super-Poincaré algebra \(\mathfrak o(p,q|\mathcal N)\) \cite{Castellani:1991et,Sati:2008eg}. These cocycles only depend on the translation \(\mathtt p_\mu\) and supertranslation \(\mathtt q_\alpha^i\) generators, which form a Lie subsuperalgebra \(\mathbb R^{p,q|\mathcal N}\) with the brackets
\begin{align}
    [\mathtt p_\mu,\mathtt p_\nu]&=0,&[\mathtt p_\mu,\mathtt q^i_\alpha]&=0,&
    [\mathtt q_\alpha^i,\mathtt q_\beta^j]=\eta^{ij}\gamma_{\alpha\beta}^\mu\mathtt p_\mu,
\end{align}
where \(\gamma_{\alpha\beta}^\mu\) are \((p,q)\)-dimensional gamma matrices and \(\eta^{ij}\) is the invariant bilinear form of the R-symmetry group. One considers the \(L_\infty\)-algebraic central extensions of \(\mathbb R^{d-1,1|\mathcal N}\) that are Lorentz-invariant --- equivalently, \(L_\infty\)-algebraic central extensions of \(\mathfrak{io}(p,q|\mathcal N)\) given by cocycles that do not involve the spacetime rotation \(\mathtt j_{\mu\nu}\) or R-symmetry \(\mathtt r^i{}_j\) generators. Such a cocycle of degree \(p+1\) then corresponds to a \(p\)-form potential \(A^{(p)}\) on spacetime.

A cocycle of degree \(p+1\) produces two kinds of couplings on a \((p-1)\)-brane worldvolume action in the Green--Schwarz formalism: a direct coupling to the \(p\)-form potential corresponding to the cocycle as well as a Wess--Zumino--Witten term for \(\kappa\)-symmetry \cite{Henneaux:1984mh,Fiorenza:2013nha}. The former is immediate: if the Green--Schwarz superembedding maps from the worldvolume \(\Sigma\) into target superspace \(\mathbb R^{p,q|\mathcal N}\) are \((x^\mu,\theta^\alpha_i)\), then the coupling to the gauge field \(A^{(p)}\) is
\begin{equation}\label{eq:term1}
    \int_\Sigma x^*A.
\end{equation}
For the latter, one starts with the invariant forms
\begin{align}
    e^\mu&\coloneqq\mathrm dx^\mu+\eta^{ij}\theta^\alpha_i\gamma_{\alpha\beta}\mathrm d\theta^\beta_j,&
    \psi^\alpha&\coloneqq\mathrm d\theta^\alpha.
\end{align}
Given a cocycle \(z(\mathtt p^\mu,\mathtt q^\alpha_i)\) (where \(\mathtt p^\mu\) and \(\mathtt q^\alpha_i\) are the translation and supertranslation generators of the Chevalley--Eilenberg algebra \(\operatorname{CE}(\mathfrak{io}(p,q|\mathcal N))\) respectively), we can consider the expression \(z(e^\mu,\psi^\alpha_i)\) where the generators \(\mathtt p^\mu\) and \(\mathtt q^\alpha_i\) have been replaced by the worldvolume fields \(e^\mu\) and \(\psi^\alpha_i\).
Then the Wess--Zumino--Witten term on the brane worldvolume is
\begin{equation}\label{eq:term2}
    \int_{\tilde\Sigma}z(e^\mu,\psi^\alpha_i),
\end{equation}
where \(\tilde\Sigma\) is a \((p+2)\)-dimensional manifold-with-boundary whose boundary is \(\Sigma\), and where \(x^\mu\) and \(\theta^\alpha_i\) have been extended to \(\tilde\Sigma\) in some fashion; the fact that \(z(e^\mu,\psi^\alpha_i)\) is closed ensures that this expression is independent of small perturbations to this extension.
This term, in turn, is crucial for ensuring \(\kappa\)-symmetry of the Green--Schwarz worldvolume action necessary for obtaining the correct number of degrees of freedom of the brane in superstring theory \cite{Simon:2011rw}.

The two terms \eqref{eq:term1} and \eqref{eq:term2} are not independent: in fact, they combine naturally as
\begin{equation}\label{eq:weil_combine}
    \int_{\tilde\Sigma}x^*\mathrm dA+z(e^\mu,\psi^\alpha_i),
\end{equation}
where the integrand may be identified as corresponding to an element of the Weil algebra of the \(L_\infty\)-algebra \(\mathfrak g\) governing supergravity \cite{Sati:2008eg}.

Furthermore, the brane bouquet constraints which branes can end on which other branes: if starting from \(\mathfrak o(p,q;\mathcal N)\) one takes a central extension
\[\mathbb R[m+1]\to\mathfrak g\to\mathfrak o(p,q;\mathcal N)\]
corresponding to an \((m-1)\)-brane and then an iterated central extension
\[\mathbb R[n+1]\to\mathfrak g'\to\mathfrak g\]
corresponding to an \((n-1)\)-brane, then the \((m-1)\)-brane can end on the \((n-1)\)-brane; for instance, the \(L_\infty\)-algebras \(\mathfrak g'\) corresponding to D-branes are obtained as central extensions of those \(L_\infty\)-algebras \(\mathfrak g\) corresponding to the fundamental strings in various superstring theories (type~IIA, type~IIB, heterotic etc.).

\subsection{Coupling to \texorpdfstring{\(p\)-}{𝑝‐}form fields}
Given a \(p\)-form field \(A^{(p)}\in\Omega^p(M)\) and the embedding map \(x\colon\Sigma\to M\), one can couple a \((p-1)\)-brane to \(A^{(p)}\) via the coupling
\begin{equation}
    S_\mathrm{int} = -m\int_\Sigma x^*A^{(p)},
\end{equation}
where \(m\) is a coupling constant and \(\Sigma\) is the \(p\)-dimensional worldvolume. For a particle, \(\Sigma\) is a line, which we can parameterise along a target-space time coordinate \(t\), so that in local coordinates on \(M\) we have \(x(t)^\mu=(1,\vec x(t))\). Expanding \(A^{(1)}=(\phi,A^{(1)}_i)\), the coupling term is then
\begin{equation}
    S_\mathrm{int} = -\int\mathrm dt\,m(\phi + \dot x^iA^{(1)}_i),
\end{equation}
so that \(-m\phi\) is the familiar coupling to the Newtonian gravitational potential \(\phi\) while \(-m\dot x^iA^{(1)}_i\) is a velocity-dependent gravitational interaction. In the case of Bargmann gravity where one sets all curvature components to zero except for that corresponding to the boost, one can always work in a gauge where \(\vec A=0\), so that one reproduces ordinary Newtonian gravity \cite{Andringa:2010it,Andringa:2013mma}.

To examine \(p\)-brane couplings to gravity for other values of \(p\), we postulate the \((p+1)\)-form curvatures (of \(p\)-form potentials) to vanish, so that we can choose the gauge choices \eqref{eq:gauge-odd} and \eqref{eq:gauge-even} for the \(p\)-form potentials.

For a \(2k\)-brane \(\Sigma\),
in adapted coordinates where \(A^{(0)}\) is taken to be the target-space as well as worldvolume time coordinate,
the coupling is then
\begin{equation}
\begin{aligned}
    S_\mathrm{int}&=-m\int x^*A^{(2k+1)}\\
    &\propto m\int\mathrm dt\,\mathrm d^{2k}\sigma\,\sqrt{\det h}\,\varepsilon^{0\alpha_1\dotso\alpha_{2k}}\partial_{\alpha_1}x^{i_1}\dotsm\partial_{\alpha_{2k}}x^{i_{2k}}\\
    &\hskip10em\times
    \left(\phi
    F^{(2)}_{i_1i_2}\dotsm F^{(2)}_{i_{2k-1}i_{2k}}
    -
    2k
    A^{(1)}_{i_1}F^{(2)}_{0i_2}\dotsm F^{(2)}_{i_{2k-1}i_{2k}}
    \right),
\end{aligned}
\end{equation}
where \(\sigma^\alpha\) are the coordinates parameterising the spatial directions of the \(2k\)-brane worldvolume and where \(h\) is the worldvolume metric, and where
\begin{equation}
F^{(2)}\coloneqq\mathrm dA^{(1)}
\end{equation}
is the field strength of \(A^{(1)}\); in components,
\begin{align}
F^{(2)}_{0i}&=\dot A^{(1)}_i
-\partial_i\phi,&
F^{(2)}_{ij}=\partial_iA^{(1)}_j-\partial_jA^{(1)}_i.
\end{align}
That is, we see that the \(2k\)-brane couples to gravity in a velocity-dependent way, depending on the \(2k\)th power of velocity; for a static \(2k\)-brane (where \(\dot x^i=0\)), the interaction term \(S_\mathrm{int}\) vanishes except when \(k=0\) (i.e.\ it is a particle).

Similarly, for a \((2k-1)\)-brane,
\begin{equation}
\begin{aligned}
    S_\mathrm{int}&=-m\int x^*A^{(2k)}\\
    &\propto m\int\mathrm dt\,\mathrm d^{2k-1}\sigma\,\sqrt{\det h}\,
    \varepsilon^{0\alpha_2\dotso\alpha_{2k}}\partial_{\alpha_2}x^{i_2}\dotsm\partial_{\alpha_{2k}}x^{i_{2k}}
    A^{(1)}_{i_2}F^{(2)}_{i_3i_4}\dotsm F^{(2)}_{i_{2k-1}j_{2k}}.
\end{aligned}
\end{equation}
Again, the coupling is velocity-dependent, and for a static (\(\dot x^i=0\)) \((2k-1)\)-brane the interaction \(S_\mathrm{int}\) vanishes.

In all these cases, if \(A^{(1)}_i=0\) (as for pure Newtonian gravity \cite{Andringa:2010it,Andringa:2013mma}), the velocity-dependent coupling to gravity vanishes: a scalar-field gravitational field does not canonically universally couple to extended objects without e.g.\ depending on the worldvolume metric.

\subsection{Wess--Zumino--Witten terms}
The cocycles we have constructed for kinematical Lie algebras depend on \(\mathtt h\) and \(\mathtt t^i_a\).
In order to construct Wess--Zumino--Witten terms following the brane-bouquet scheme,
we can introduce brane embedding maps
\begin{equation}(t,x^i_a)\colon \Sigma\to \mathbb R^{1+2d}\end{equation}
that involve a doubling of the number of spatial coordinates (but not the time coordinate).\footnote{That is, our `doubled spacetime' is simply taken to be the homogeneous space \(\mathfrak g/\mathfrak o(d)\cong\mathbb R^{1+2d}\), assumed to be flat here for simplicity. For a discussion of more sophisticated ways of constructing spacetimes from a kinematical Lie algebra \(\mathfrak g\), see \cite{Geracie:2015dea}.}
This doubling is superficially reminiscent of the doubling of coordinates in double field theory \cite{Aldazabal:2013sca,Hohm:2013bwa,Berman:2020tqn} and doubled sigma models \cite{Tseytlin:1990va,Tseytlin:1990nb,Hull:2004in,Berman:2007xn,Copland:2011wx,DeAngelis:2013wba} --- specifically, a situation where all spatial directions are compactified (but not time) and thus doubled.
From the standpoint of kinematical Lie algebras, this doubling is natural insofar as the \cref{def:kinematical_lie_algebra} of a kinematical Lie algebra does not distinguish between the boost generators \(\mathtt t_i^1\) and the spatial translation generators \(\mathtt t_i^2\), so that it makes sense to treat them on an equal footing.

In the present case, two questions arise: (1) Are the Wess--Zumino--Witten terms necessary, as in the case of superstring theory, given that there is no supersymmetry and no analogue of \(\kappa\) symmetry? (2) Given the doubled spatial coordinates, can we recover ordinary physics with undoubled degrees of freedom?
For (1), the necessity of the Wess--Zumino--Witten terms comes from the fact that they naturally combine with couplings to the \(p\)-form potentials as described in \eqref{eq:weil_combine}, so that from the perspective of \(L_\infty\)-algebras it seems unnatural to drop one but not the other. As to (2), the answer is that in some (but not all) cases, as long as kinetic terms are only present for half of the coordinates \(x^i_a\) (say, the ones corresponding to spatial translations rather than to boosts), the remaining half of the coordinates become auxiliary. This is because the Wess--Zumino--Witten terms are at most linear in the derivatives of a specific \(x^i_a\), so that without a term quadratic in the derivatives, their Euler--Lagrange equations of motion become algebraic.

For example, let us consider the case of the usual Bargmann central extension  of the Galilean algebra.
Consider a particle-brane with worldline \(\Sigma=\partial\tilde\Sigma\) that is the boundary of the surface \(\tilde\Sigma\). Consider the embedding map
\begin{equation}
    (t,x^i,v^i)\colon \tilde M\to\mathbb R^{1+2d},
\end{equation}
where -- anticipating later interpretation -- we have labelled the doubled spatial coordinates as \(x^i\) and \(v^i\) rather than \(x^i_a\).
For the Galilean algebra, from the Lie brackets we read off the invariant forms
\begin{align}
    e^i_1 &= \mathrm dv^i,&
    e^i_2 &= \mathrm dx^i + v^i \,\mathrm dt - t \,\mathrm dv^i.
\end{align}
For the Bargmann central extension \(\mathtt t^i_1\mathtt t^j_2\delta_{ij}\),
the Wess--Zumino--Witten term is
\begin{equation}
    \int_{\tilde\Sigma}
    e^i_1\wedge e^j_2\delta_{ij}
    =
    \int_\Sigma\left(
    v^i\,\mathrm dx_i
    +
    t v^i\,\mathrm dv_i\right).
\end{equation}
If we add a kinetic term for \(x^i\), then the gauge-fixed worldline Polyakov action is
\begin{equation}
\begin{aligned}
    S&=\int_\Sigma
    \left(
    \frac12\dot x^i\dot x_i
    +
    v^i\,\dot x_i
    +
    t v^i\,\dot v^i\right)\,\mathrm dt\\
    &=\int_\Sigma
    \left(
    \frac12\dot x^i\dot x_i
    +
    v^i\,\dot x_i
    -\frac12v_iv^i\right)\,\mathrm dt,
\end{aligned}
\end{equation}
so that the equation of motion for \(v^i\) reads
\begin{equation}
    \dot x^i = v^i.
\end{equation}
Integrating \(v^i\) out, we find
\begin{equation}
    S = \int_\Sigma\dot x^i\dot x_i,
\end{equation}
which is equivalent to the ordinary Polyakov action of a particle save for an overall factor. Thus we recover the un-doubled physics of a particle, and the equations of motion force \(v^i\) to be interpreted as the velocity of the particle.

However, this is not always the case. If we had instead unwisely put a kinetic term for \(v^i\) instead of \(x^i\), then \(x^i\) would act as a Lagrange multiplier enforcing \(\dot v^i=0\), which leads to trivial physics.
Similarly, in the case of the static algebra \(\mathfrak{newt}(d;(\begin{smallmatrix}0&0\\0&0\end{smallmatrix}))\) (where \(x^i\) and \(v^i\) are on a completely equal footing),
we obtain a trivial physics regardless of which kinetic terms we choose.

\section*{Acknowledgements}
Hyungrok Kim thanks Julian Matteo Kupka\textsuperscript{\orcidlink{0009-0003-3701-9709}} for pointers to literature and helpful comments, as well as Leron Borsten\textsuperscript{\orcidlink{0000-0001-9008-7725}}, Dimitri Kanakaris Decavel\textsuperscript{\orcidlink{0009-0001-7716-851X}}, and Christian Sämann\textsuperscript{\orcidlink{0000-0002-5273-3359}} for helpful discussion.

\newcommand\cyrillic[1]{\fontfamily{Domitian-TOsF}\selectfont \foreignlanguage{russian}{#1}}
\bibliographystyle{unsrturl}
\bibliography{biblio}
\end{document}